\documentclass[prc,preprint,nofootinbib,amsmath,amssymb,aps]{revtex4-1}
\usepackage{graphicx}
\usepackage{dcolumn}
\usepackage[usenames]{color}
\usepackage{float}
\usepackage{placeins}
\usepackage{marvosym}

\newcolumntype{d}[1]{D{.}{.}{#1}}

\newcommand{\simge}{\hspace*{0.2em}\raisebox{0.5ex}{$>$}
     \hspace{-0.8em}\raisebox{-0.3em}{$\sim$}\hspace*{0.2em}}

\newcommand{\beq}{\begin{equation}}
\newcommand{\eeq}{\end{equation}}
\newcommand{\bea}{\begin{eqnarray}}
\newcommand{\eea}{\end{eqnarray}}

\begin{document}
\title{Renormalization of One-Pion Exchange in Higher Partial Waves\\
	    in Chiral Effective Field Theory for Antinucleon-Nucleon System}

\author{Daren Zhou}
\affiliation{School of Electrical and Computer Engineering, Nanfang College, Guangzhou, Guangdong 510970, China}

\date{\today}
\vspace{3em}

\begin{abstract}
The renormalization of iterated one-pion exchange (OPE) has been studied in Chiral Effective Field Theory ($\chi$EFT) for the antinucleon-nucleon ($\overline{N}\!N$) scattering in some partial waves (Phys. Rev. C 105, 054005 (2022)). We go further for the other higher partial waves but with total angular momenta $J\leq 3$ in this paper. Contact interactions are represented by a complex spherical well in coordinate space.
Changing the radius of the spherical well means changing the cutoff.
We check the cutoff dependence of the phase shifts, inelasticities, and mixing angles for the partial waves, and show that contact interactions are needed at leading order in channels where the singular tensor potentials of OPE are attractive.
Results are compared with the energy-dependent partial-wave analysis of $\overline{N}\!N$ scattering data.
Comparisons between our conclusions and applications of $\chi$EFT to the nucleon-nucleon system are discussed as well.
\end{abstract}

\maketitle

\section{Introduction} \label{sec:intro}

Renormalization~\cite{Gell-Mann1954,Wilson1971a,Wilson1971b,Wilson1971c,Wilson1975} is an important subject in quantum field theory and which was developed to deal with ultraviolet divergences. After renormalization, physical observables shall be finite and independent of `cutoff', i.e.,
renormalization-group (RG) invariant.
Traditionally, it was thought that a theory is renormalizable if all of its coupling constants in the Lagrangian have nonnegative mass dimensionality, otherwise it is nonrenormalizable and thus not that useful.
However, the modern view of renormalizability is different~\cite{Weinberg:1995,Lepage:1997,Lepage:2005,Zee:2010}, which is based on the idea of Effective Field Theory (EFT).

Quantum Chromodynamics (QCD) is regarded as a correct theory of strong interaction, and which is part of the standard model of particle physics. QCD can be solved perturbatively at high energies because of asymptotic freedom. However, QCD is nonperturbative at low energies where most of nuclear interactions happen.
One way to solve the problems in low energy range is using lattice QCD, another way is using effective theories. Chiral Effective Field Theory ($\chi$EFT) is an effective theory of QCD at low energies,
where the degrees of freedom are no longer quarks and gluons but baryons and pions instead.
$\chi$EFT applies when the characteristic momentum $Q$ in a process is well below the QCD nonperturbative scale $M_{\rm QCD}\sim 1$ GeV.
From the point of view of an effective theory, one can add as many terms as one wants in an effective Lagrangian
as long as they obey the required symmetries. Of course, these terms must be ordered in a proper way based on some power counting.
For an effective theory at low energies, the details of its high energy physics are not shown explicitly or are unknown, and which effects are embedded in interaction strengths or ``low-energy constants'' (LECs).
Once the Lagrangian is obtained, physical observables can be calculated in a systematic expansion in powers of
$Q/M_{\rm QCD}$~(abbreviated as $Q$ hereafter if there is no confusion)~\cite{Weinberg:1978kz}.
This way is obviously model independent. Ultraviolet divergences can be
absorbed by the LECs via order-by-order renormalization.

Weinberg's seminal articles~\cite{Weinberg:1990rz,Weinberg:1991um} have been inspiring a lots of applications of EFTs to a variety of nuclear, hypernuclear, and antinuclear systems.
Since the first implementation~\cite{Ordonez:1992xp,Ordonez:1994,Ordonez:1995rz} in this direction, significant improvements have been made, for reviews, see for instance Refs.~\cite{Machleidt:2011zz,Epelbaum:2012vx,Hammer:2019poc,Hergert:2020bxy}.
In Refs.\cite{Kang:2013uia,Dai:2017ont}, Weinberg's proposal was extended to antinucleon-nucleon ($\overline{N}\!N$) system.
Here we discuss the renormalization of one-pion exchange (OPE) in $\chi$EFT for $\overline{N}\!N$ scattering at low energies and which is an extension of Ref.~\cite{Zhou:2022}.

Renormalization was studied enormously (most in the framework of EFTs) for $N\!N$ system \cite{Kaplan:1996xu,Frederico:1999ps,Beane:2001bc,Eiras:2003,Valderrama2004,Timoteo2005,Valderrama2005a,Valderrama2005b,Nogga:2005hy,
Birse:2006um,Valderrama2006,PavonValderrama:2006uj,Birse:2007,Long:2007vp,yang2008,Higa2008,Yang:2009kx,Yang:2009pn,Birse:2009my,
Epelbaum2009,Valderrama:2011ei,Harada2011,Timoteo2011,Long:2011qx,PavonValderrama:2011fcz,Long:2011xw,Epelbaum2013,Epelbaum2018,
Epelbaum2020,Epelbaum2021,Gasparyan:2021,Liu:2022cfd,Gasparyan:2022},
and still there are some disagreements.
It was known that there are inconsistencies in Weinberg's power counting (WPC), i.e., observables obtained with Weinberg's approach are not RG invariant \cite{Kaplan:1996xu,Beane:2001bc,Nogga:2005hy,PavonValderrama:2006uj,Kaplan:1998a,Kaplan:1998b,Savage:1998vh}.
(For different views, see for instance Refs.~\cite{Epelbaum2013,Epelbaum2018,Epelbaum2020,Epelbaum2021,Gasparyan:2021,Gasparyan:2022}.)
The reason for this is that WPC is based on naive dimensional analysis (NDA) \cite{Manohar:1983md}, and which has its origin in application of naturalness~\cite{tHooft:1979rat,Veltman:1980mj} to perturbative series~\cite{vanKolck:2020plz}
while $N\!N$ has nonperturbative character at LO.
It is well known that nonperturbative renormalization can be very different from the perturbative one~\cite{vanKolck:2020llt}.
Once a channel contains an attractive OPE tensor force which has $\sim -r^{-3}$ singular part, an extra
counterterm is needed than expected by NDA when the OPE is treated nonperturbatively \cite{Beane:2000wh,Nogga:2005hy,PavonValderrama:2007nu}.
In other words, the LECs are in fact promoted by the nonperturbative RG running.

In this paper, we extend our previous study on the renormalization of the $\overline{N}\!N$ scattering amplitudes~\cite{Zhou:2022} to other higher partial waves with total angular momentum $J\leq 3$.
The OPE will be treated nonperturbatively as before.
Again, we use the results of the Groningen energy-dependent partial-wave analysis (PWA) \cite{Zhou:2012ui} as a kind of base.
The Groningen PWA gives a quite good description of the data for elastic $\overline{p}p\rightarrow\overline{p}p$ and charge-exchange $\overline{p}p\rightarrow\overline{n}n$ scattering below 925 MeV/$c$ antiproton laboratory momentum.
In the previous paper~\cite{Zhou:2022}, we have studied the renormalization of OPE in the uncoupled $S$, $P$,
and the coupled $S$-$D$ and $P$-$F$ partial waves.
Here, we focus on the uncoupled $D$, $F$, and the coupled $D$-$G$ parital waves.
The analysis is similar to those of Ref.~\cite{Nogga:2005hy,Zhou:2022},
and we use the same the boundary conditions as in the Groningen PWA,
with a coordinate-space regulator~\cite{Scaldeferri:1996nx,Beane:2000wh,Beane:2001bc}.
At a certain energy, the LEC is fitted to the phase shift and inelasticity of the PWA, with phase shifts and inelasticities at other energies and mixing angles being postdictions of the EFT at LO. We work in the isospin basis, since isospin breaking is a higher-order effect. In future studies, we will try to treat OPE
perturbatively in some $\overline{N}\!N$ partial waves \cite{Chen:2010an,Chen:2011yu}
and include other higher order interactions.

The organization of this paper is the following. In Sec.~\ref{sec:OPE}, we describe the OPE potential and the contact interactions, and identify the partial waves which have attractive singular tensor forces and thus need additional counterterms than expected by NDA. In Sec.~\ref{sec:Strategy}, the renormalization strategy is explained and the results are shown. The conclusions and outlooks are given in Sec.~\ref{sec:Conclusions}.

\section{One-pion exchange and contact interactions} \label{sec:OPE}

Since $\chi$EFT is a low energy EFT of QCD, the degrees of freedom are nucleon ($N$) and antinucleon ($N^c$) now in an antinuclear system, and for processes involving momenta $Q$ on the order of the pion mass,
the three pions ($\pi^a$, with $a=1,\,2,\,3$) must also be
included explicitly as pseudo-Goldstone bosons from the spontaneous breaking of chiral symmetry.
The interactions are constrained by the approximate chiral symmetry and other symmetries of the QCD dynamics.
OPE arises from the first few terms in the nonrelativistic chiral Lagrangian density~\cite{Oosterhof:2019dlo}
\begin{eqnarray}
\mathcal{L}_{\overline{N}\!N} &=&
N^\dagger \left(i \partial_0+\frac{\nabla^2}{2m_N}\right) N
+ {N^c}^\dagger\left(i \partial_0+\frac{\nabla^2}{2m_N}\right)N^c
-\frac{1}{2}\pi^a\left(\partial^2+m_\pi^2\right)\pi^a
\nonumber\\
&&
+ \,\frac{g_A}{2 f_\pi}\left({N^\dagger}\vec{\sigma}\tau^a N
+ {{N^c}^\dagger}{\vec\sigma}\tau^{aT} N^c\right)\cdot\vec{\nabla}\pi^a
+\dots\,, \label{eq:lagran}
\end{eqnarray}
where $m_N\simeq 940$ MeV is the (anti)nucleon mass, $m_\pi\simeq 138$ MeV is the pion mass,
$g_A\simeq 1.29$ is the axial-vector coupling constant, $f_\pi\simeq 92.4$ MeV is the pion decay constant
and $\tau^a$ ($\vec{\sigma}$) are the Pauli matrices for isospin (spin).
The contact interactions are buried in the dots.
The electromagnetic terms are not included since we are studying the strong interaction here.

The energy transfer between (anti)nucleons is of order $Q^2/m_N$ after renormalization
and thus the pion propagator can be taken as approximately static.
At LO, i.e., at the order of $Q^0$,
the OPE potential in coordinate space for the $\overline{N}\!N$ system can be obtained from Eq. (\ref{eq:lagran}) and which is
\begin{equation}
V_\pi = -\frac{g_A^2 m_\pi^3}{16\pi f_\pi^2}  \,\boldsymbol{\tau}_1\cdot\boldsymbol{\tau}_2\,
\bigl[v_S(m_\pi r) \,\vec{\sigma}_{1}\cdot\vec{\sigma}_{2}
+ v_T(m_\pi r)\,S_{12}\bigr] \, ,
\label{OPE}
\end{equation}
where
$S_{12}=3\,\vec{\sigma}_{1}\cdot\hat{r}\,\vec{\sigma}_{2}\cdot\hat{r} -\vec{\sigma}_{1}\cdot\vec{\sigma}_{2}$
is the tensor operator, and
\begin{subequations}
	\begin{eqnarray}
	v_S(x) & = & \frac{e^{-x}}{3x}~, \\
	v_T(x) & = & \left(1+x+\frac{x^2}{3}\right) \frac{e^{-x}}{x^3} \ .
	\end{eqnarray}
\end{subequations}
At next-to-leading order (NLO), i.e., at the order of $Q$, there is no correction because of Lorentz invariance and parity invariance~\cite{Weinberg:1991um,Ordonez:1992xp}.
At next-to-next-to-leading order (${\text N}^2$LO), i.e., at the order of $Q^2$, chiral-symmetry-breaking corrections represent the Goldberger-Treiman discrepancy, which increases the strength of static OPE \cite{vanKolck:1996rm,vanKolck:1997fu}, while chiral-symmetric corrections to the pion-(anti)nucleon interaction account for recoil \cite{Ordonez:1992xp}. Two-pion-exchange (TPE) also appears at this order~\cite{Ordonez:1992xp}.
Eq.~\eqref{OPE} has the opposite sign as OPE for $N\!N$ (cf., for example, Ref. \cite{Nogga:2005hy}),
because the $G$ parity of the pions is $-1$.

For higher partial waves, the perturbative treatment of the OPE potential may be proper and enough
at low energies, but since there is yet no clear boundary between nonperturbation and perturbation~\cite{vanKolck:2020llt}, we will leave this for further study, and will continue to treat OPE potential nonperturbatively in this paper as we have done in Ref.~\cite{Zhou:2022}. The nonperturbative results presented here can also be used as a comparison for the future perturbative results, and which may also be helpful in identifying the boundary between nonperturbation and perturbation if it exists.

As one can see in Eq.~\eqref{OPE}, the OPE potential is singular and which diverges as $r^{-3}$ and thus needs to be regularized. The regulator we use here is the same as the one used in the Groningen PWA~\cite{Zhou:2012ui}, which follows the procedure of the Nijmegen PWAs of Refs.~\cite{Timmermans:1990tz,Timmermans:1994pg,Timmermans:1995xb}.
Where, the short-range interaction and long-range interaction is separated at a boundary $b=1.2$ fm.
The long-range potential consists of the electromagnetic interaction, OPE, and TPE
\cite{Rentmeester:1999vw,Rentmeester:2003mf}.
At $r=b$ a boundary condition is chosen that, for convenience, corresponds to a complex spherical well which is independent of energy, depends on the spin and isospin of the partial wave.
The partial-wave Schr\"odinger equation is solved for the coupled $\overline{p}p$ and $\overline{n}n$ channels with a long-range potential with $r>b$.
In this paper, the long-range potential only contains OPE
and the partial-wave Schr\"odinger equation can be solved in a similar way as in Ref.~\cite{Zhou:2012ui}.
Solving the Schr\"odinger equation with OPE potential is equivalent to iterating OPE to all orders.
The boundary radius $b$, however, is not fixed at $1.2$ fm and can vary now. The relation between the ultraviolet cutoff $\Lambda$ and the boundary radius $b$ is $\Lambda\equiv 1/b$.

In $\chi$EFT, the short-range QCD dynamics are embedded in the contact interactions. For $\overline{N}\!N$, they are complex instead of real in $N\!N$ case.
The complex character is responsible for the complicated annihilation processes.
Mesons generated in these processes with energies on the order of $2m_N$ cannot be accounted for as explicit degrees of freedom in the EFT.
One example is the annihilation diagram into one pion, related to OPE by crossing, which would be present in a relativistic theory.
There are also annihilation states containing soft pions which give rise to long-range effects, but their contributions to elastic scattering are suppressed by a factor of $(Q/4\pi f_\pi)^2$, as for other irreducible loops.
Annihilation also generates contributions to the real parts of the LECs in addition to their imaginary parts, which makes the LECs of $\overline{N}\!N$ different from those of $N\!N$.
Like any other EFT LECs, these contact interactions must appear at orders no higher than where they are needed to remove arbitrary regulator dependence. The actual values of the real and imaginary parts will be determined by fitting to the empirical PWA ``data''.

The contact interactions, which are (derivatives of) Dirac delta functions in coordinate space, are smeared with a spherical well~\cite{Scaldeferri:1996nx,Beane:2000wh,Beane:2001bc}
resembling the one used in Ref.~\cite{Zhou:2012ui}. Schematically, for a channel $c$,
\begin{equation}
   C_{c} \, \mathcal{O}_{c}\delta^{(3)}(r)
   \rightarrow
   \frac{3C_{c}}{4\pi b^3}\,\theta(b-r)\,\mathcal{P}_{c}
   \equiv \left(V_c+iW_c\right)\, \theta(b-r) \, \mathcal{P}_{c} \ ,
\label{eqn:counter}
\end{equation}
where $C_c$ is a complex LEC, $\mathcal{O}_{c}$ is a combination of derivatives, $\mathcal{P}_c$ is the projection operator on channel $c$, and the real short-range parameters $V_c$ and $W_c$ from different channels are independent. Therefore, the regulator is proportional to Heaviside function.
For a different way of defining the counterterms of the annihilation, see Refs.~\cite{Kang:2013uia, Dai:2017ont}.

Since we are considering uncoupled $D$, $F$ and coupled $D$-$G$ partial waves here, the short-range
interactions are expected by NDA to appear at even higher orders than those of the lower partial waves considered in Ref.~\cite{Zhou:2022}.
However, whether a short-range interaction is needed at LO in a certain channel
hinges on the tensor part of OPE being attractive,
as having been discussed for the $N\!N$ case in Ref.~\cite{Nogga:2005hy,PavonValderrama:2006uj}
and for the $\overline{N}\!N$ case in Ref.~\cite{Zhou:2022}.
Therefore, here the short-range interactions in the partial waves
whose tensor part of OPE being attractive seems need to be promoted to LO as well,
and we will check this.

A partial wave is denoted by $^{2I+1\;2S+1}L_J$, where $I$ ($S$) is the total isospin (spin) and $L$ ($J$) is the orbital (total) angular momentum.
Because the Pauli principle does not apply to the $\overline{N}\!N$ system, the total isospin is independent from other quantum numbers,
which is unlike in the $N\!N$ system where $I+S+L$ must be odd.
Moreover, due to annihilation
there are four times as many phase parameters (phase shifts, inelasticities, and mixing angles) compared to $np$ scattering, viz. 20 phase parameters for each value of $J\neq0$ \cite{Timmermans:1995xb,Tim84}.
For uncoupled partial waves, the $S$ matrix is just a complex number with two real parameters,
\begin{equation}
S = \eta\,e^{2i\delta} \ ,
\end{equation}
where $\delta$ is the phase shift and $\eta$ ($0\leq\eta\leq1$) is the inelasticity due to
annihilation. For the coupled spin-triplet partial waves with $L=J\mp1$ ($J\ge1$) the $2\times 2$ $S$ matrix is parameterized
as~\cite{Timmermans:1994pg,Zhou:2012ui}
\begin{equation}
S^J = \exp(i\bar{\delta}) \,
\exp(i\varepsilon_{J}\sigma_x) \,\,
H^J  \, \exp(i\varepsilon_{J}\sigma_x) \,
\exp(i\bar{\delta})~,
\end{equation}
where $\bar{\delta}$ is a diagonal matrix with real entries
$\delta_{J-1,J}$ and $\delta_{J+1,J}$, and $\varepsilon_{J}$
is the mixing angle.
The matrix $H^J$ parameterizes the inelasticities. It is written as
\begin{equation}
H^J = \exp(-i\omega_J\sigma_y) \,
\left( \begin{array}{cc}
\eta_{J-1,J} &       0        \\
0        &   \eta_{J+1,J}
\end{array} \right) \,
\exp(i\omega_J\sigma_y)~,
\label{Eq:Klarsfeld}
\end{equation}
where $\eta_{J-1,J}$ and $\eta_{J+1,J}$ are the inelasticities
($0 \leq\eta_{J\mp1,J}\leq 1$) and
$\omega_J$ is the mixing angle for inelasticity.
The $S$ matrix for the coupled partial waves is thus written in terms of six real parameters.

Since the sign of the singular tensor force is a crucial point for the renormalization property of a certain partial wave, let us figure them out.
The matrix elements of the tensor force have similar properties as for $N\!N$.
From Eq.~\eqref{OPE}, one can see that the sign of the singular tensor force depends on the matrix elements of $\boldsymbol{\tau}_1\cdot\boldsymbol{\tau}_2$ and $S_{12}$.
The spin-singlet channels do not have singular long-range forces since the matrix elements of $S_{12}$ between those channels vanish.
These spin-singlet channels are $^1D_2$ and $^1F_3$ waves in this paper.
For the spin-triplet channels, the matrix elements of $S_{12}$ do not vanish.
The operator $\boldsymbol{\tau}_1\cdot\boldsymbol{\tau}_2=2I(I+1)-3=-3,\,+1$ for $I=0,\,1$, respectively, therefore, for the spin-triplet uncoupled channels, which always makes one channel attractive and its isospin partner repulsive.
These spin-triplet uncoupled channels are $^3D_2$ and $^3F_3$ waves in this paper.
We tabulate in Table~\ref{tab:signs} the values of the matrix element of $-\boldsymbol{\tau}_1\cdot\boldsymbol{\tau}_2 \,S_{12}$ in these partial waves.
The eigenvalues of $S_{12}$ are $2$ and $-4$, regardless of $J$, therefore, for the spin-triplet coupled channels, always one eigenchannel is attractive and the other repulsive.
These spin-triplet coupled channels are $^3D_3$-$^3G_3$ waves in this paper.
On account of $\boldsymbol{\tau}_1\cdot\boldsymbol{\tau}_2$, the isoscalar eigenchannel is most attractive.

In summary, we have the partial waves $^{11}D_2$, $^{31}D_2$, $^{11}F_3$ and $^{31}F_3$ which have no tensor force, and the partial waves $^{13}D_2$ and $^{13}F_3$ which have repulsive tensor force,
and the partial waves $^{33}D_2$ and $^{33}F_3$ which have attractive tensor force,
and the partial waves $^{13}D_3$-$^{13}G_3$ and $^{33}D_3$-$^{33}G_3$ which
always one eigenchannel is attractive and the other repulsive.
\begin{table}[tb]
	\centering
	\caption{Values of $-\boldsymbol{\tau}_1\cdot\boldsymbol{\tau}_2 \,S_{12}$
		in the spin-triplet uncoupled $D$ and $F$ waves.}
	\tabcolsep=2.1em
	\renewcommand{\arraystretch}{0.9}
	\begin{tabular}{ccccccccc}
		\hline
		\hline
		Partial wave & $^{13}D_2$ & $^{33}D_2$ & $^{13}F_3$ &   $^{33}F_3$ \\
		\hline
		$-\langle\boldsymbol{\tau}_1\cdot\boldsymbol{\tau}_2 \,S_{12}\rangle$
		& $+6$ & $-2$ & $+6$ & $-2$ \\
		\hline
		\hline
	\end{tabular}
	\label{tab:signs}
\end{table}
In the next section we confirm numerically that partial waves having attractive tensor force need additional LECs than expected by NDA and the others do not need,
and compare the phase shifts, inelasticities and mixing angles they yield with the empirical PWA values.

\section{Renormalization and results} \label{sec:Strategy}

The strategy is the same as in Ref.~\cite{Zhou:2022}
where the renormalization of OPE for the lower partial waves were discussed.
Firstly, we check the cutoff dependence of the phase shifts and mixing angles before renormalization.
This can be done by solving the partial-wave Schr\"odinger equation with $V_c=W_c=0~{\rm fm}^{-1}$,
since the other parameters of OPE are known.
(When $W_c=0~{\rm fm}^{-1}$, inelasticities are trivial, i.e., $\eta_{J\mp 1,J}=1$, and thus their mixing angles $\omega_J$ are not well determined by Eq.~(\ref{Eq:Klarsfeld}) and thus all of them are not shown before renormalization.)
We show the results at various representative laboratory energies
($T_{\rm lab} = 10,\,50,\,100$ MeV).
Secondly, we adjust the strength of short-range spherical well as function of $b$
for each partial wave where cutoff dependence is found.
Here $V_c$ and $W_c$ play the role of counterterms, whose cutoff dependence ensures that physical observables be cutoff independent within error bars. We determine $V_c$ and $W_c$ by fitting to the phase shift and inelasticity of the Groningen PWA~\cite{Zhou:2012ui} at some energy.
The fit results are not significantly sensitive to the choice of fitting energy, which is taken to be $T_{\rm lab}=20$ MeV.
Thirdly, after verifying that cutoff independence is achieved,
we take a cutoff value $\Lambda \sim M_{\rm QCD}$ (specifically, $\Lambda= 5$ fm$^{-1}$) and compare phase shifts, inelasticities, and mixing angles as functions of the laboratory momentum $p_{\rm lab}$
(between $0$ and $450$ MeV/c) with the PWA. The PWA values we show here contain more points but are consistent with the values in Tables VIII and IX of Ref.~\cite{Zhou:2012ui}, which assume isospin symmetry.

In the following, we split the analysis between spin-singlet and triplet channels, since the former ones have no tensor force and the latter ones have, and the tensor force is the determining factor for the renormalization of OPE.

\subsection{Spin-singlet channels} \label{singlets}

\begin{figure}[b]
	\centering
	\includegraphics[width=0.45\textwidth]{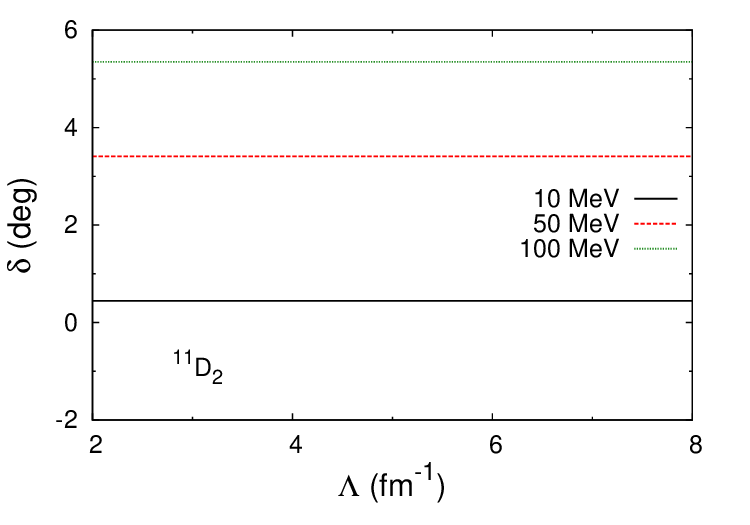} \hspace{2em}
	\includegraphics[width=0.45\textwidth]{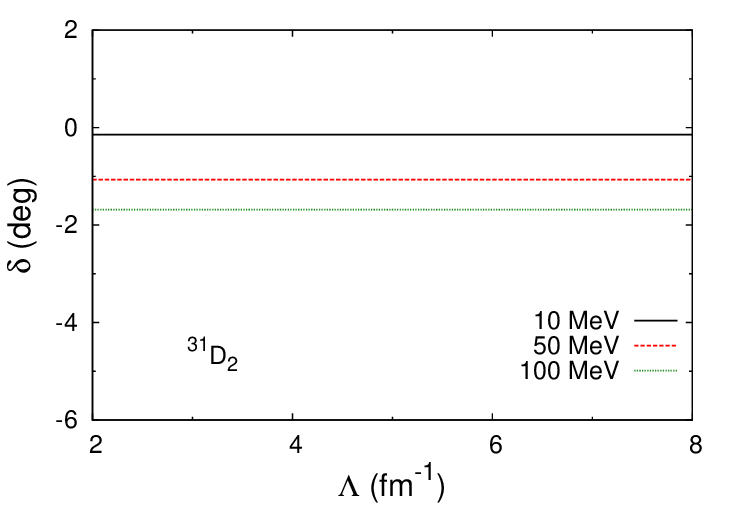}\\	
	\includegraphics[width=0.45\textwidth]{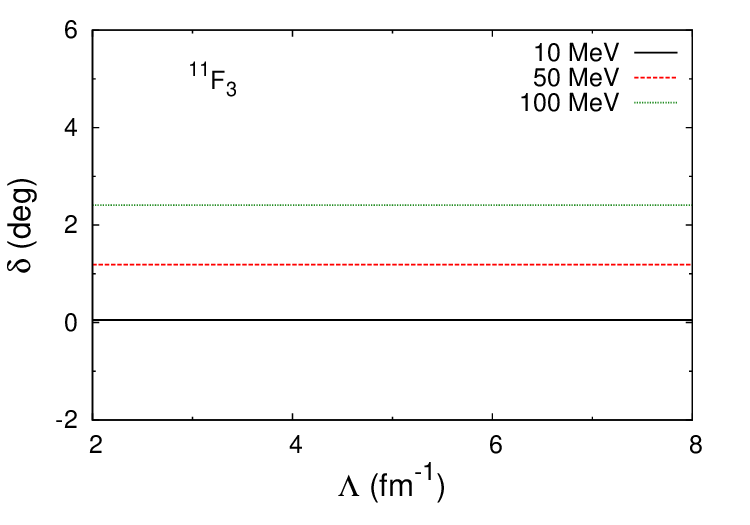} \hspace{2em}
	\includegraphics[width=0.45\textwidth]{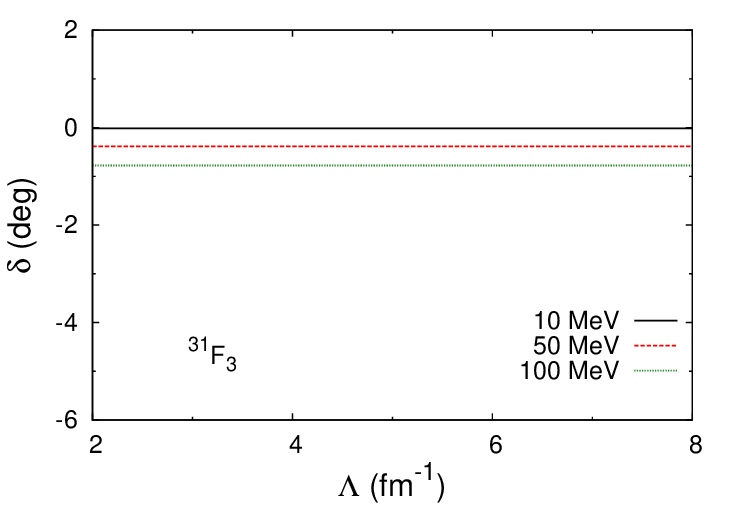}\\
	\caption{\label{SingletPhase_L}{Cutoff dependence of the phase shifts in the spin-singlet $D$ and $F$ waves at the laboratory energies of 10 MeV (black solid line), 50 MeV (red dashed line), and 100 MeV (green dotted line), for $V_c=W_c=0~\text{fm}^{-1}$.}}
\end{figure}

\begin{figure}[H]
	\centering
	\includegraphics[width=0.45\textwidth]{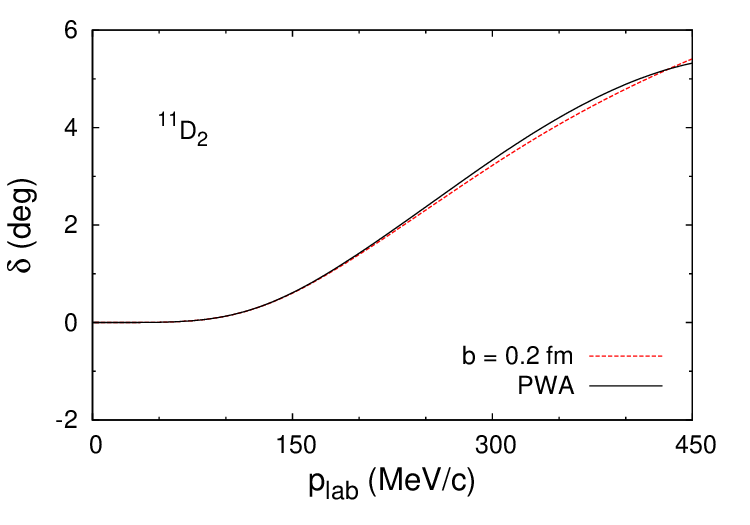} \hspace{2em}	
	\includegraphics[width=0.45\textwidth]{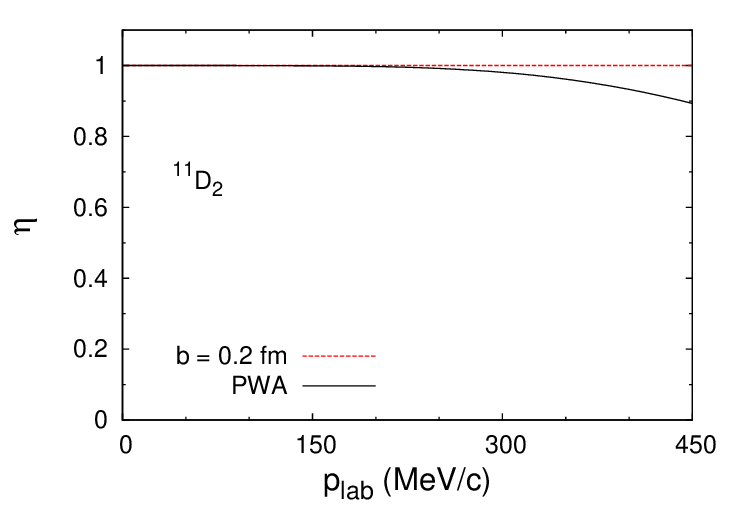} \\
	\includegraphics[width=0.45\textwidth]{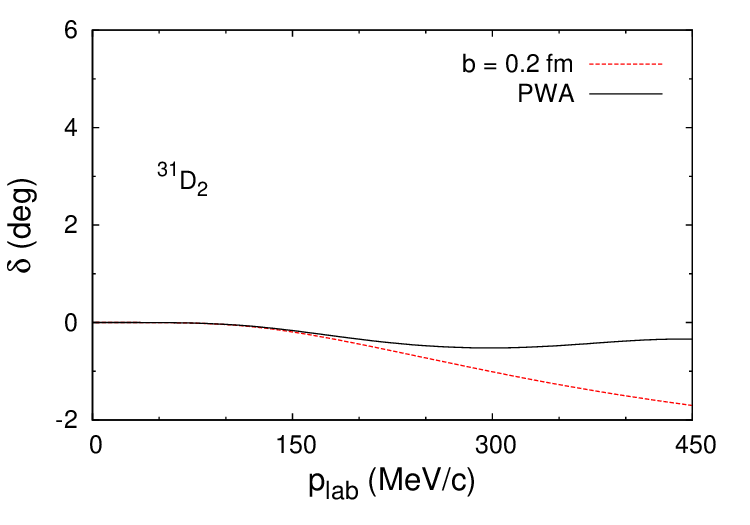} \hspace{2em}	
    \includegraphics[width=0.45\textwidth]{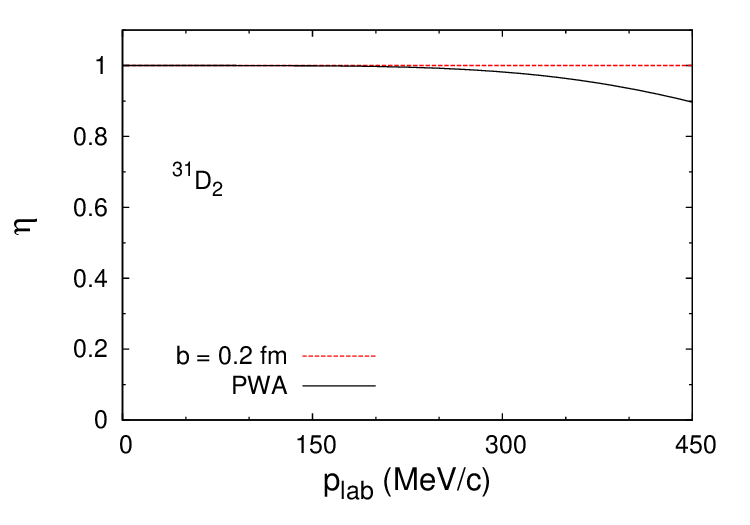} \\
	\includegraphics[width=0.45\textwidth]{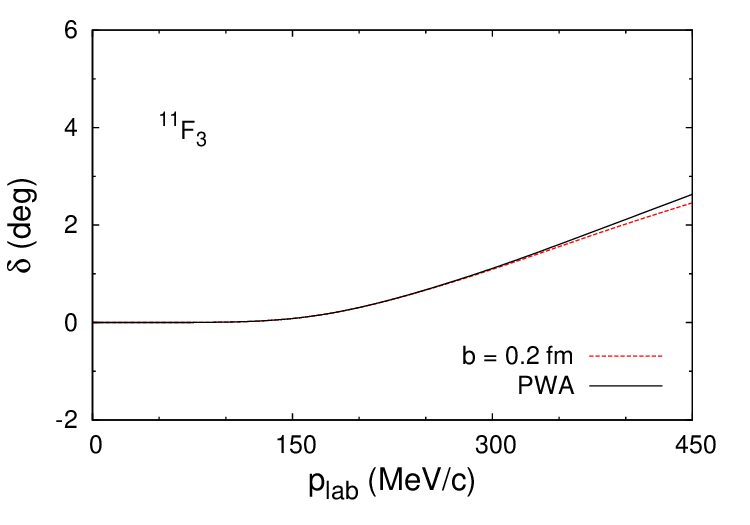} \hspace{2em}	
    \includegraphics[width=0.45\textwidth]{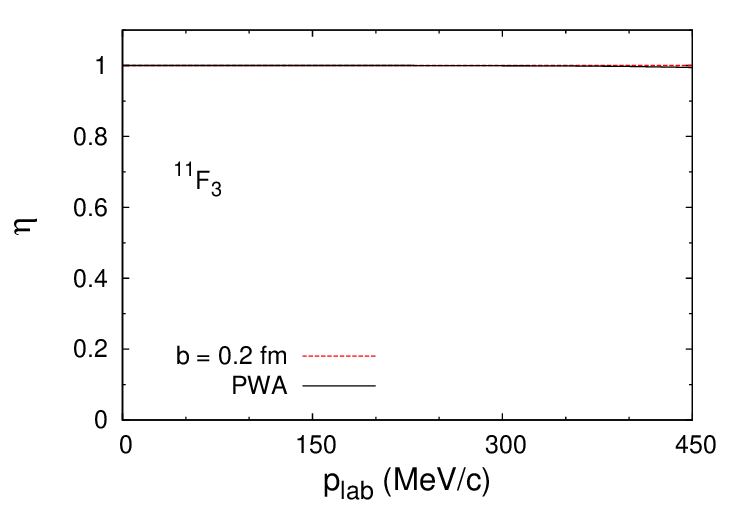} \\
    \includegraphics[width=0.45\textwidth]{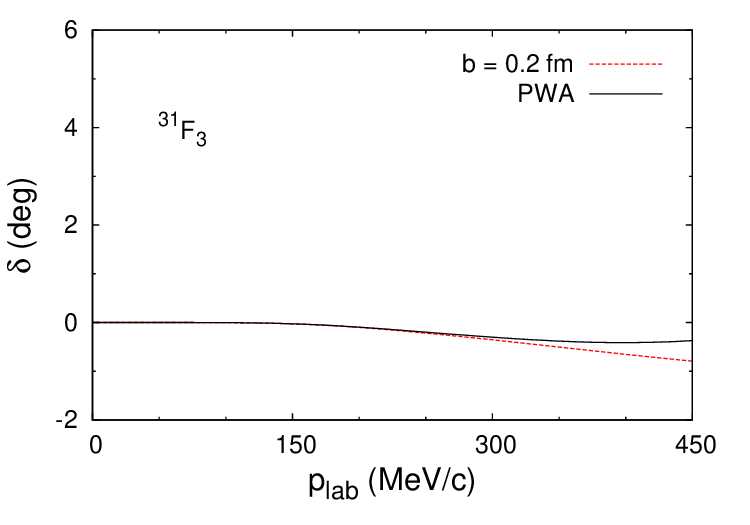} \hspace{2em}	
    \includegraphics[width=0.45\textwidth]{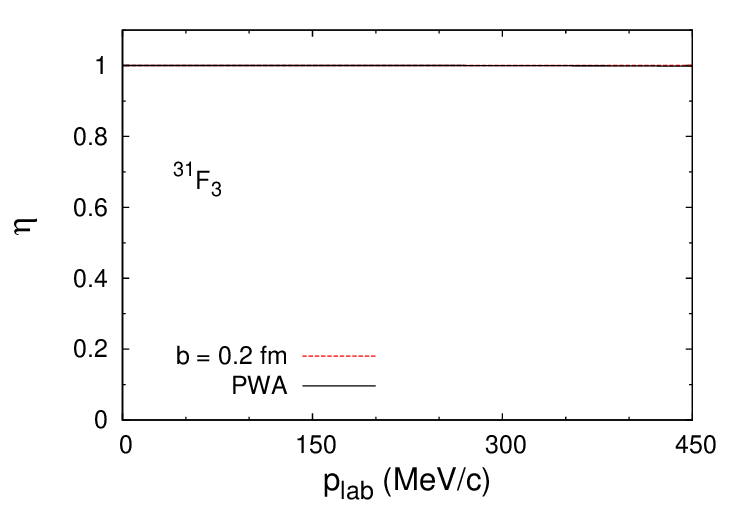} \\
	\caption{\label{SingletPhase_plab}{Phase shifts (left panels) and inelasticities (right panels) of the spin-singlet $D$ and $F$ waves against laboratory momentum. The (red) dashed lines are from iterated one-pion exchange for $b=0.2$ fm and $V_c=W_c=0~\text{fm}^{-1}$, while (black) solid lines are the results of the PWA~\cite{Zhou:2012ui}.}}
\end{figure}

The spin-singlet channels have the $^{11}D_2$, $^{31}D_2$, $^{11}F_3$ and $^{31}F_3$ partial waves.
According to NDA, the counterterms of the $^{1}D_2$ waves start to appear on the order of $Q^4$,
and the counterterms of the $^{1}F_3$ waves start to appear on the order of $Q^6$.
They do not need to be promoted to LO since the OPE potential has no tensor forces in these channels.
In Fig.~\ref{SingletPhase_L}, one can see that the phase shifts are obviously cutoff independent
in the cutoff range between $2$ ${\rm fm}^{-1}$ and $8$ ${\rm fm}^{-1}$.
All phase shifts approach finite values as the cutoff increases.
The isospin-singlet phase shifts are attractive and relatively large while the isospin-triplet ones are repulsive and relatively small due to the factor
$-\boldsymbol{\tau}_1\!\cdot\!\boldsymbol{\tau}_2\,\vec{\sigma}_{1}\!\cdot\!\vec{\sigma}_{2}$.
The $\eta$s are all equal to $1$ since $W_c=0~{\rm fm}^{-1}$ here, and thus are not shown in the plots.
In Fig.~\ref{SingletPhase_plab}, the phase shifts and inelasticities
against laboratory momentum and the comparison with the PWA values are shown,
where $V_c$ and $W_c$ are from Table~\ref{tab:Uncp_vw}.
The results agree with the PWA values well in the energy range considered except for
the phase shift of $^{31}D_2$,
which means that in this wave high order contributions might needed in order to make a better agreement with the PWA.
These agreements with the PWA
are comparable with those obtained in the $N\!N$ case~\cite{Nogga:2005hy}.
Thus, a satisfactory description of the spin-singlet channels is obtained at
low energies with a LO consisting of iterated OPE and no contact interaction.

\begin{table}[h]
	\centering
	\caption{Values of the real ($V_c$) and imaginary ($W_c$) components of the short-range potential at $b=0.2$ fm for the uncoupled $D$ and $F$ partial waves. These of $^{33}D_2$ and $^{33}F_3$ waves are obtained by fitting to the PWA ``data"~\cite{Zhou:2012ui} at $T_{\rm lab}=20$ MeV, while these of the other waves are set to be $0~{\rm fm}^{-1}$ by hand.}
	\tabcolsep=0.6em
	\renewcommand{\arraystretch}{0.9}
	\begin{tabular}{cd{4.2}d{4.2}d{4.2}d{4.2}d{4.2}d{4.2}d{4.2}d{4.2}}
		\hline
		\hline
		Partial wave &\multicolumn{1}{c}{$^{11}D_2$}&\multicolumn{1}{c}{$^{31}D_2$}&\multicolumn{1}{c}{$^{11}F_3$} & \multicolumn{1}{c}{$^{31}F_3$} &\multicolumn{1}{c}{$^{13}D_2$}&\multicolumn{1}{c}{$^{33}D_2$}&\multicolumn{1}{c}{$^{13}F_3$} & \multicolumn{1}{c}{$^{33}F_3$}\\
		\hline
		$V_c$ (fm$^{-1}$)     & 0 & 0 & 0 & 0 & 0 & -86.96 & 0 & -162.73\\
		$W_c$ (fm$^{-1}$)     & 0 & 0 & 0 & 0 & 0 & -0.05 & 0 & -0.00\\
		\hline
		\hline
	\end{tabular}
	\label{tab:Uncp_vw}
\end{table}

\subsection{Spin-triplet channels} \label{triplets}

\begin{figure}[b]
	\centering
	\includegraphics[width=0.45\textwidth]{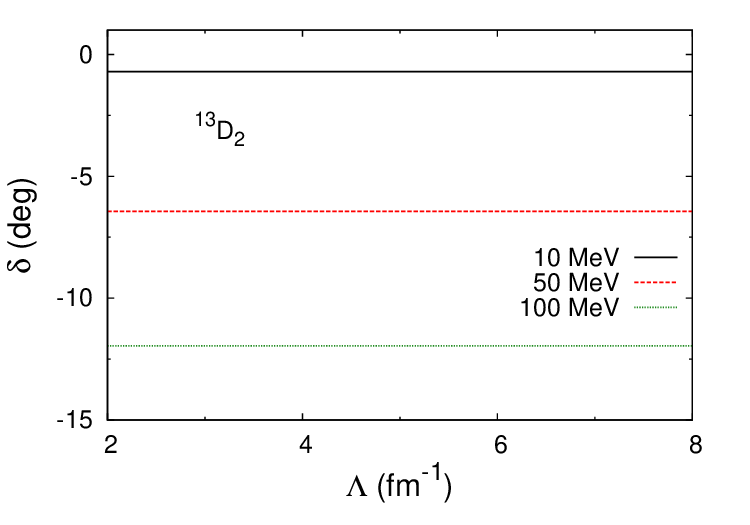} \hspace{2em}	
	\includegraphics[width=0.45\textwidth]{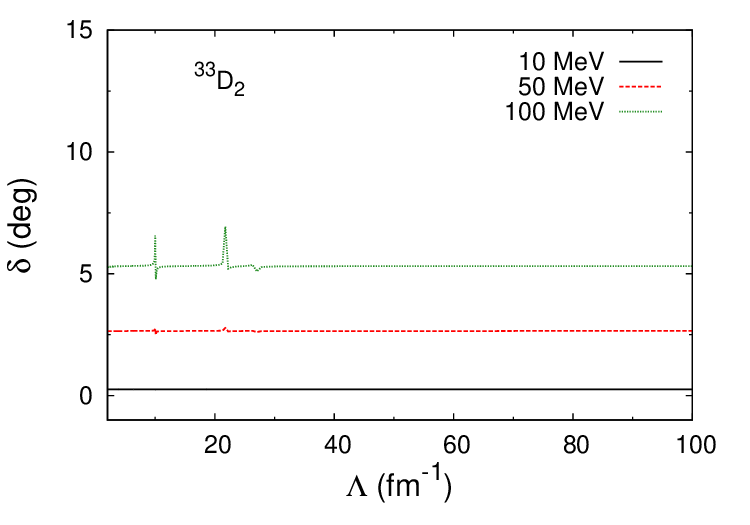} \\
	\includegraphics[width=0.45\textwidth]{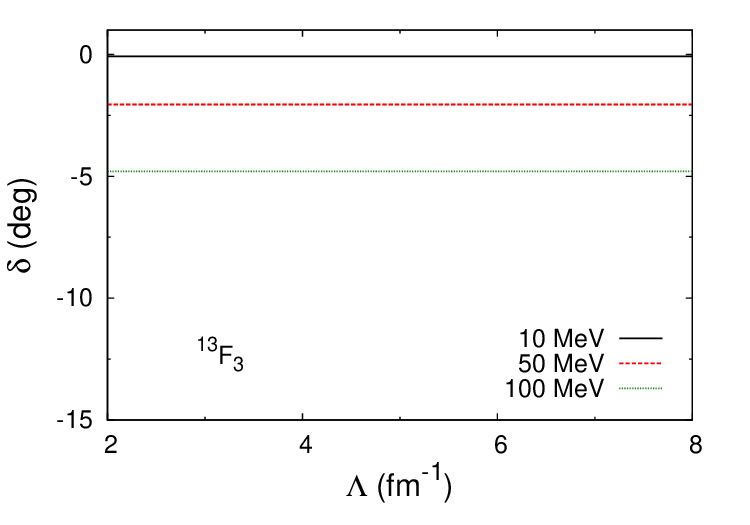} \hspace{2em}
	\includegraphics[width=0.45\textwidth]{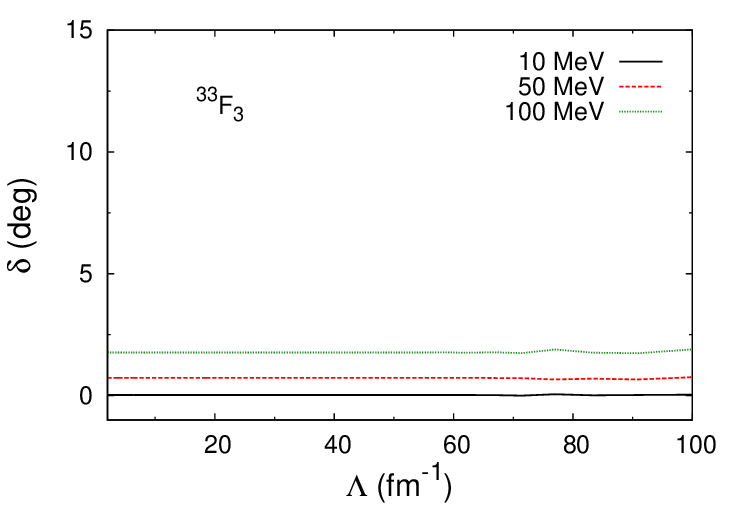} \\
	\caption{\label{TripletUncpPhases_L}{Cutoff dependence of the phase shifts in the spin-triplet uncoupled $D$ and $F$ waves at the laboratory energies of 10 MeV (black solid line), 50 MeV (red dashed line), and 100 MeV (green dotted line), for $V_c=W_c=0~\text{fm}^{-1}$.}}
\end{figure}

\begin{figure}[tb]
	\centering
	\includegraphics[width=0.45\textwidth]{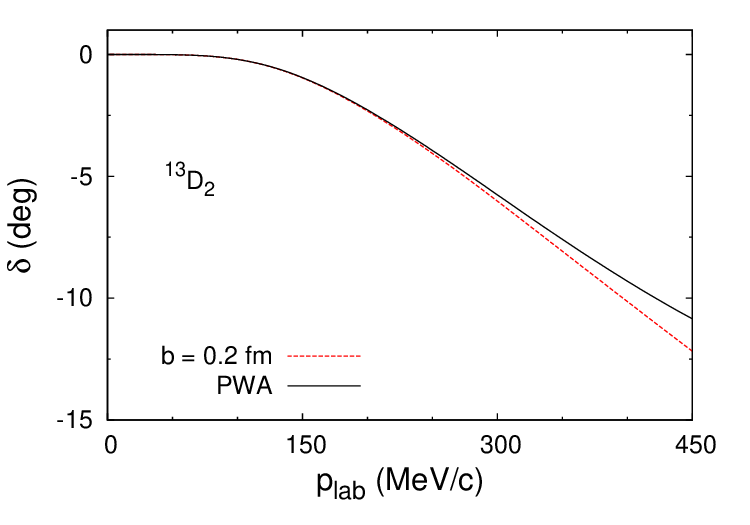} \hspace{2em}	
	\includegraphics[width=0.45\textwidth]{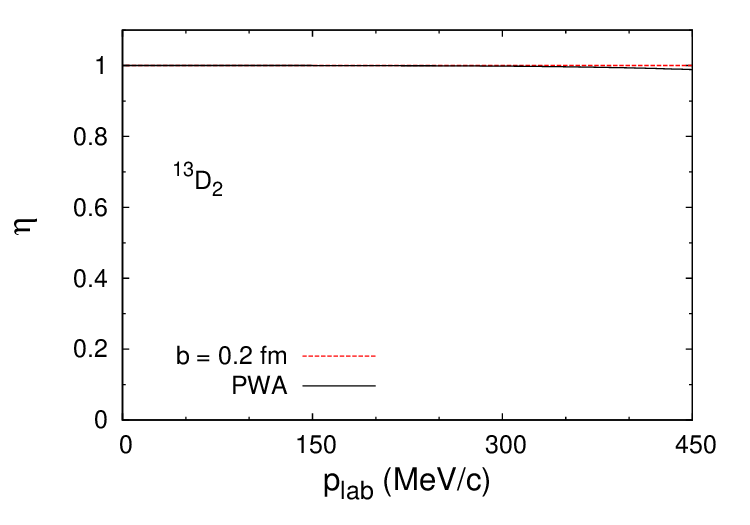} \\
	\includegraphics[width=0.45\textwidth]{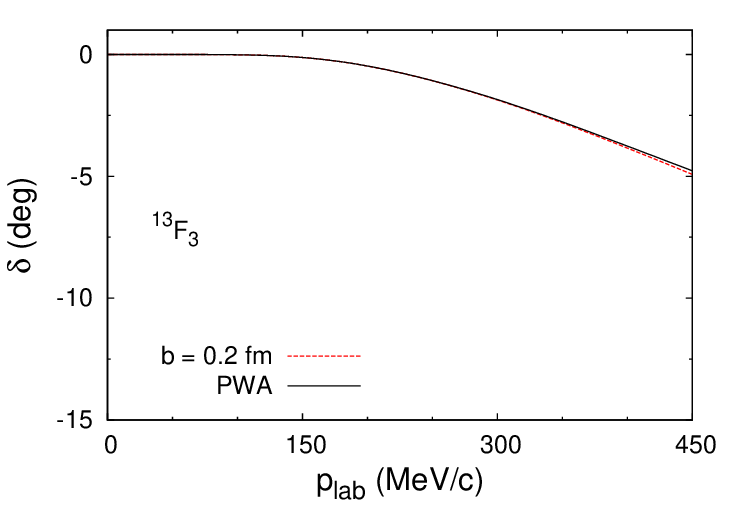} \hspace{2em}	
	\includegraphics[width=0.45\textwidth]{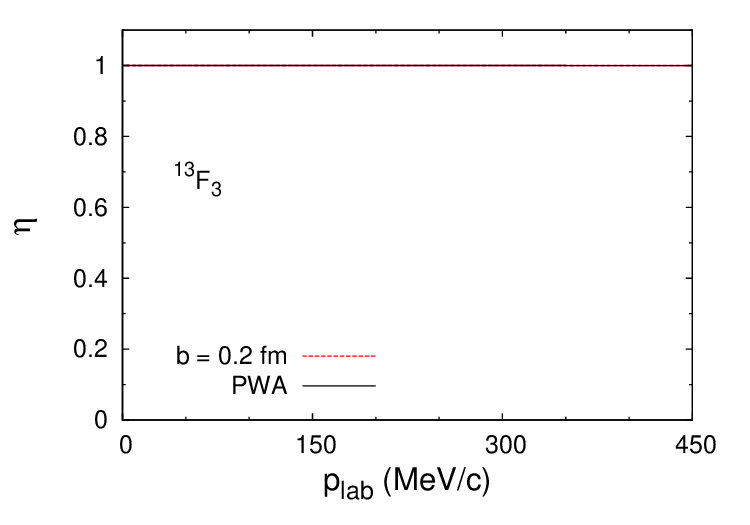} \\
	\caption{\label{TripletUncpPhases_plab1}{Phase shifts (left panels) and inelasticities (right panels) of the spin-triplet uncoupled $^{13}D_2$ and $^{13}F_3$ waves against laboratory momentum. The (red) dashed lines are from iterated one-pion exchange for $b=0.2$ fm and $V_c=W_c=0~\text{fm}^{-1}$, while (black) solid lines are the results of the PWA~\cite{Zhou:2012ui}.}}
\end{figure}

Now, we turn to the spin-triplet channels. First, we focus on the uncoupled partial waves, and the signs of the tensor forces of these waves can be seen from Table~\ref{tab:signs}.
In Fig.~\ref{TripletUncpPhases_L}, the cutoff dependence of the phase shifts of these $D$ and $F$ waves are shown. The phase shifts of the $^{13}D_2$ and $^{13}F_3$ waves are obviously
cutoff independent and they converge to finite values as the cutoff increases,
because these channels have no attractive singular tensor force.
One can find that the phase shift of the $^{33}D_2$ wave is obviously cutoff dependent though only in some limited ranges when the cutoff is high enough since this channel has an attractive singular tensor force.
The $^{33}F_3$ wave also has an attractive singular tensor force.
However, the phase shift of the $^{33}F_3$ wave only shows very tiny cutoff dependence
at the laboratory kinetic energy $T_{\rm lab}=100$ MeV,
and when the cutoff is high enough.
This is because of that the contact interaction of this wave starts to appear
on the order of $Q^6$ according NDA.
It should be expected that the cutoff dependence in this wave
would be more obvious when go to higher energies.
The different strengths of cutoff dependence of these two waves
are also due to their centrifugal barriers via orbital angular momenta.
In Fig.~\ref{TripletUncpPhases_L}, the cutoff ranges of the isospin-singlet and the isospin-triplet waves are different, because we want to see the cutoff dependence in the isospin-triplet waves since which is expected from the experience in the lower partial waves
of $\overline{N}\!N$ system~\cite{Zhou:2022}.

\begin{figure}[tb]
	\centering
	\includegraphics[width=0.45\textwidth]{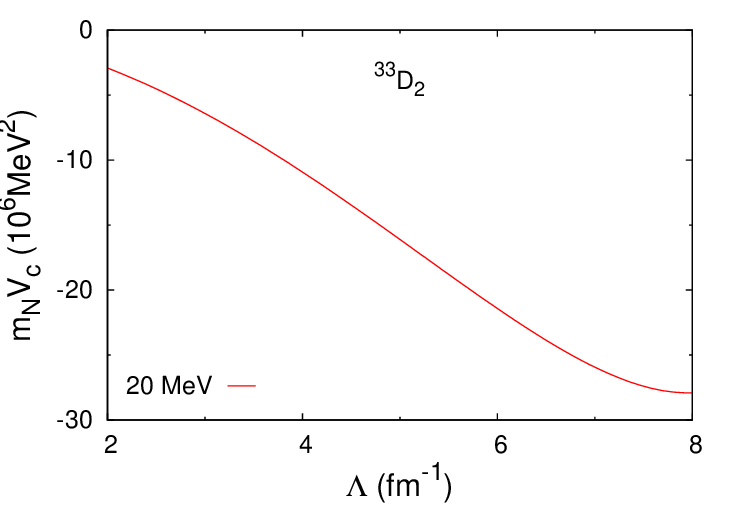} \hspace{2em}
	\includegraphics[width=0.45\textwidth]{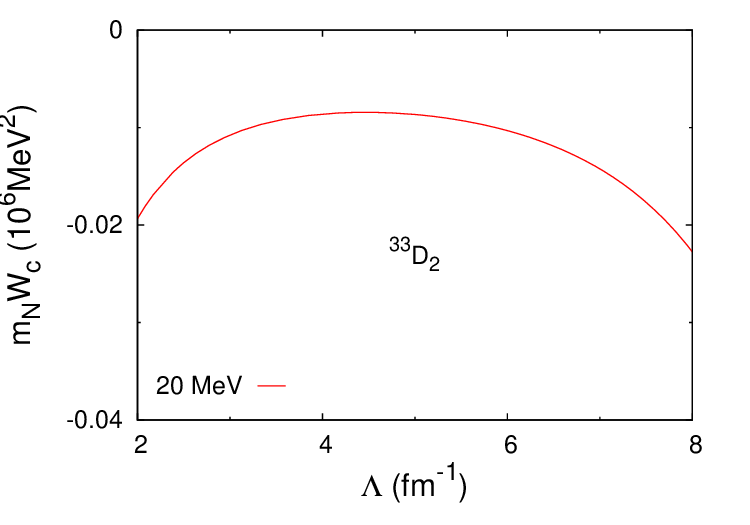} \\
	\includegraphics[width=0.45\textwidth]{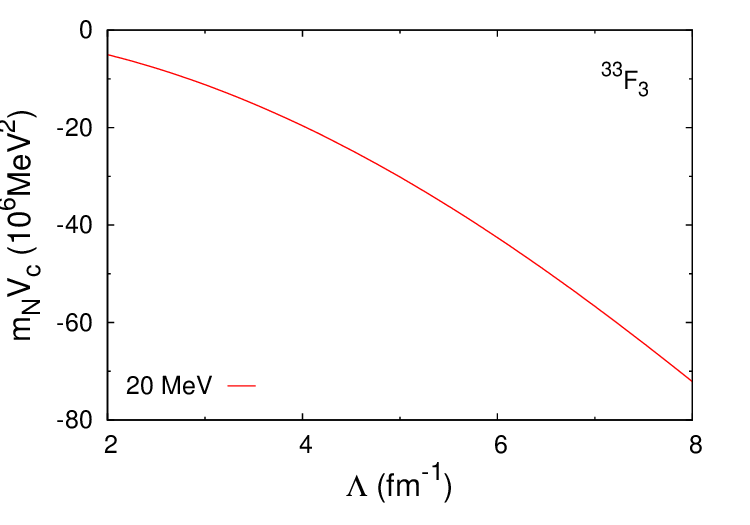} \hspace{2em}
	\includegraphics[width=0.45\textwidth]{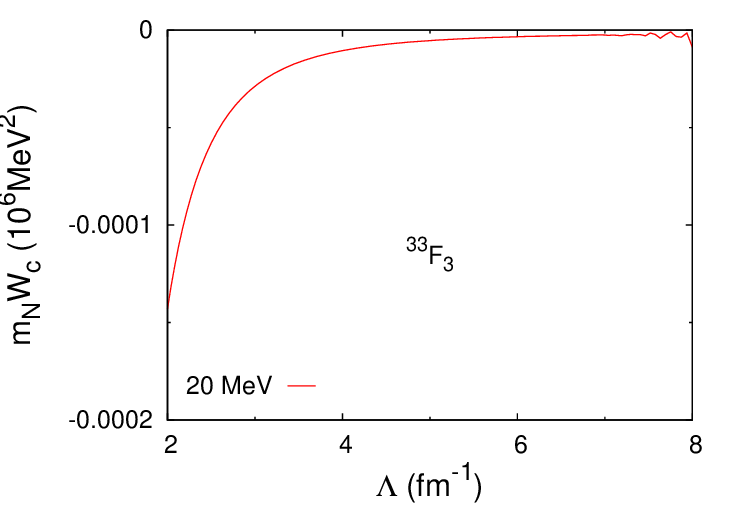} \\
	\caption{\label{TripletUncpCount_L}{Cutoff dependence of $m_N V_c$ (left panels) and $m_N W_c$ (right panels) for the spin-triplet uncoupled $^{33}D_2$ and $^{33}F_3$ waves. The PWA phase shifts and inelasticities are fitted at $T_{\rm lab}=20$ MeV.}}
\end{figure}

We have seen that the phase shifts of the $^{13}D_2$ and $^{13}F_3$ waves are cutoff independent
since these waves have repulsive tensor forces, and thus they do not need counterterms at LO
which is expected by NDA,
and therefore we set $V_c=W_c=0~{\rm fm}^{-1}$ in these channels.
In Fig.~\ref{TripletUncpPhases_plab1},
the phase shifts and inelasticities against laboratory momentum of the $^{13}D_2$ and $^{13}F_3$ waves are shown, and where the cutoff of the iterated OPE is taken to be $b=0.2~{\rm fm}$ or $\Lambda=5~{\rm fm}^{-1}$.
The results agree with the PWA values very well in the energy range considered.
Again, these agreements with the PWA values
are comparable with those obtained in the $N\!N$ case~\cite{Nogga:2005hy}.
The $\eta$s are almost $1$ in these two waves,
which means that there are almost no annihilation in these channels.

\begin{figure}[tb]
	\centering
	\includegraphics[width=0.45\textwidth]{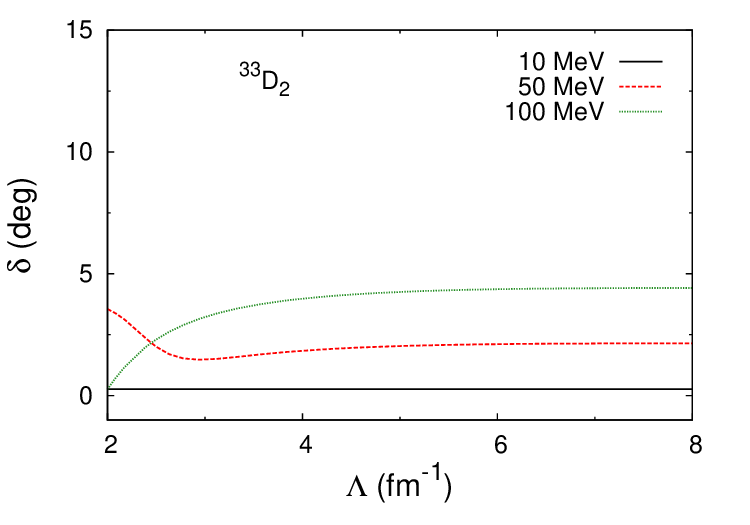} \hspace{2em}
	\includegraphics[width=0.45\textwidth]{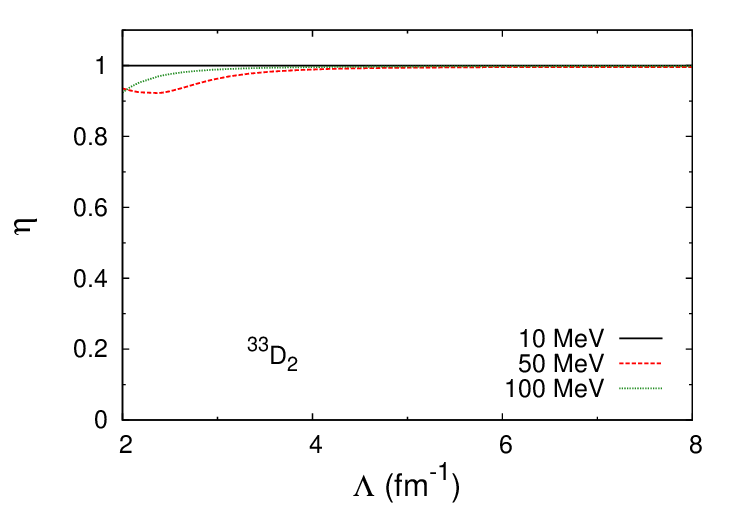}\\
	\includegraphics[width=0.45\textwidth]{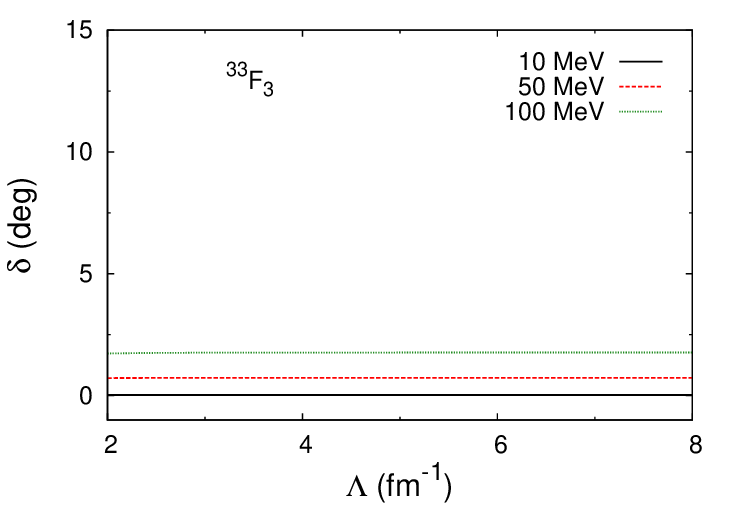} \hspace{2em}
	\includegraphics[width=0.45\textwidth]{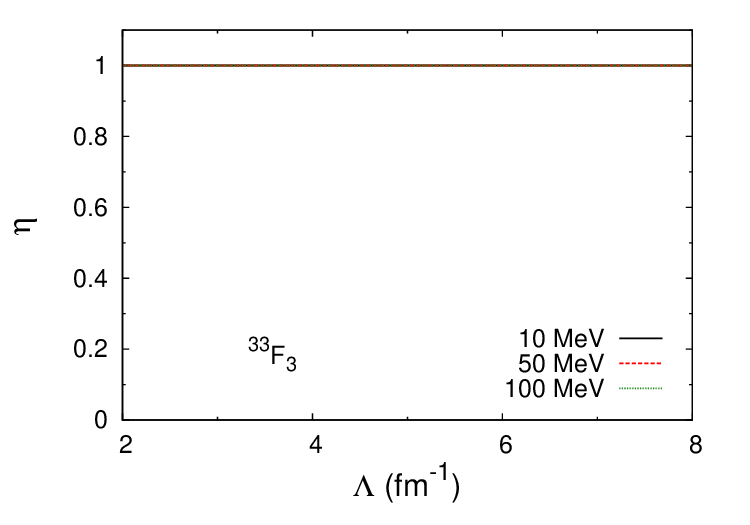} \\
	\caption{\label{TripletUncpPhases_LC}{Residual cutoff dependence of the phase shifts and inelasticities in the spin-triplet uncoupled $^{33}D_2$ and $^{33}F_3$ waves at the laboratory energies of 10 MeV (black solid line), 50 MeV (red dashed line), and 100 MeV (green dotted line), for $V_c$ and $W_c$ in Fig. \ref{TripletUncpCount_L}.}}
\end{figure}

\begin{figure}[tb]
	\centering
	\includegraphics[width=0.45\textwidth]{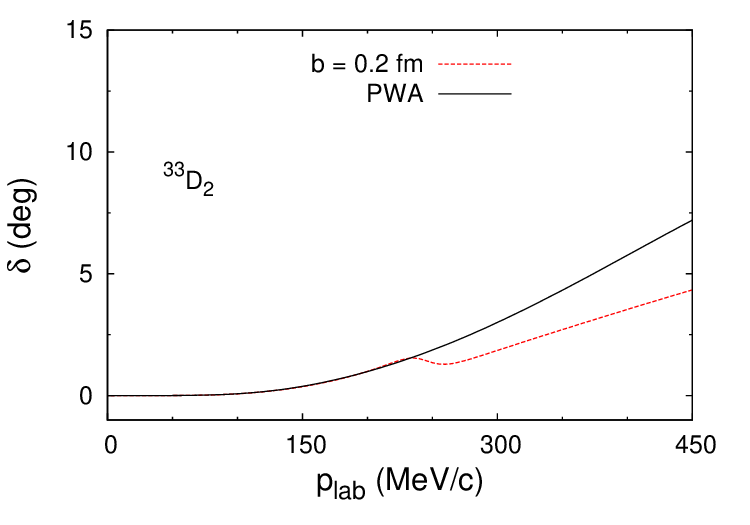} \hspace{2em}
	\includegraphics[width=0.45\textwidth]{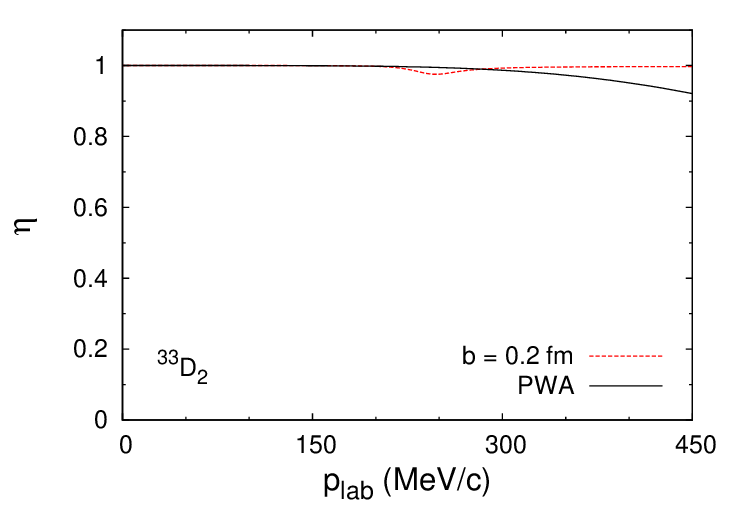} \\
	\includegraphics[width=0.45\textwidth]{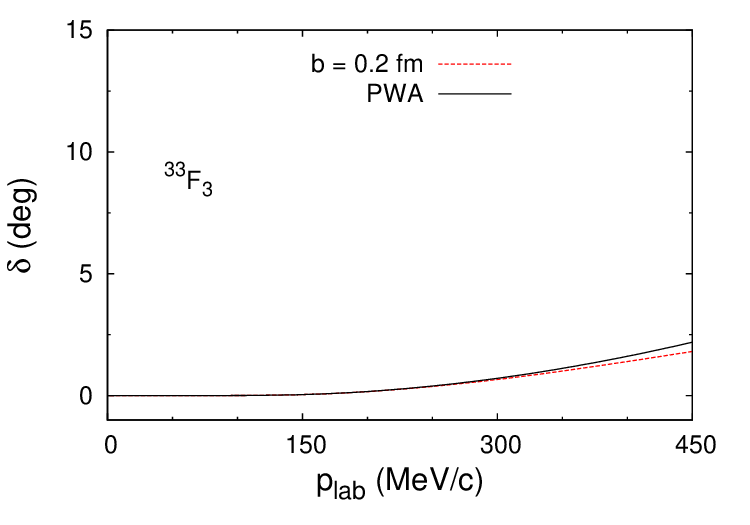} \hspace{2em}
	\includegraphics[width=0.45\textwidth]{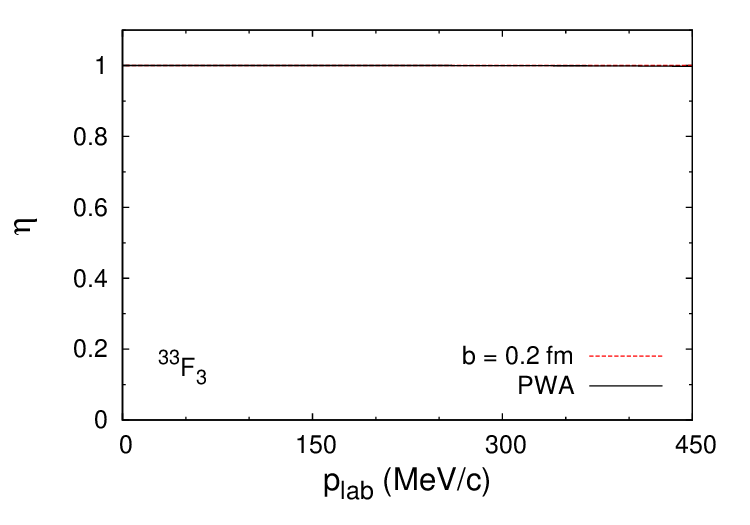}\\
	\caption{\label{TripletUncpPhases_plab2}{Phase shifts (left panels) and inelasticities (right panels) of the spin-triplet uncoupled $^{33}D_2$ and $^{33}F_3$ waves against laboratory momentum. The (red) dashed lines are from iterated one-pion exchange for $b=0.2$ fm and $V_c$, $W_c$ from Table~\ref{tab:Uncp_vw},
	while (black) solid lines are the results of the PWA~\cite{Zhou:2012ui}.}}
\end{figure}

In the other two spin-triplet uncoupled waves, $^{33}D_2$ and $^{33}F_3$, where the OPE tensor force
is singular and attractive, the counterterms are expected though the cutoff dependence is not very obvious in
the $^{33}F_3$ channel at energies considered here.
In Fig.~\ref{TripletUncpCount_L},
the running of the corresponding LECs $V_c$ and $W_c$ is shown for the $^{33}D_2$ and $^{33}F_3$ waves.
The values of these LECs are obtained by fitting to the corresponding phase shifts and inelasticities of the PWA at laboratory kinetic energy $T_{\rm lab}=20$ MeV.
The magnitudes of $V_c$ are much greater than the magnitudes of $W_c$, which can also be seen in the other partial waves in Ref.~\cite{Zhou:2022}, and the relative difference between $V_c$ and $W_c$ becomes larger as the orbital angular momentum $L$ increases. The tiny ``ripples'' on the tail part of the curve in the plot of $W_c$ of the $^{33}F_3$ wave are due to that
where the values of $W_c$ are too small and thus the quality of fitting is not as good as those in the lower cutoff range, say $\Lambda<7~{\rm fm}^{-1}$.
The residual cutoff dependence of the phase shifts and inelasticities of the $^{33}D_2$ and $^{33}F_3$ waves is shown in Fig.~\ref{TripletUncpPhases_LC}. We see that, after renormalization,
the phase shifts and inelasticities are cutoff independent and converge to finite values
as increasing cutoff.
Their phase shifts and inelasticities as functions of the laboratory momentum are plotted in Fig.~\ref{TripletUncpPhases_plab2} for $b = 0.2$ fm,
where the LECs take the values given in Table~\ref{tab:Uncp_vw}.
The results of the $^{33}F_3$ wave agree with the PWA values very well in the energy range considered.
The results of the $^{33}D_2$ wave agree with the PWA values very well
when $p_{\rm lab}\lesssim230$ MeV/c.
However, there are dents in both the curves of phase shift and inelasticity
in the $^{33}D_2$ wave when $p_{\rm lab}\simeq250$ MeV/c,
which might need higher order contributions to compensate,
or which might be due to an unnecessary iteration of OPE.
The dents will move to right as the cutoff $\Lambda$ decreases.
Of course, the actual reason for the dents need further investigation.
No this kind of dent has been found in the $N\!N$ case~\cite{Nogga:2005hy}
or in the lower partial waves in the $\overline{N}\!N$ case~\cite{Zhou:2022}.

\FloatBarrier

\begin{figure}[tb]
	\centering
	\includegraphics[width=0.45\textwidth]{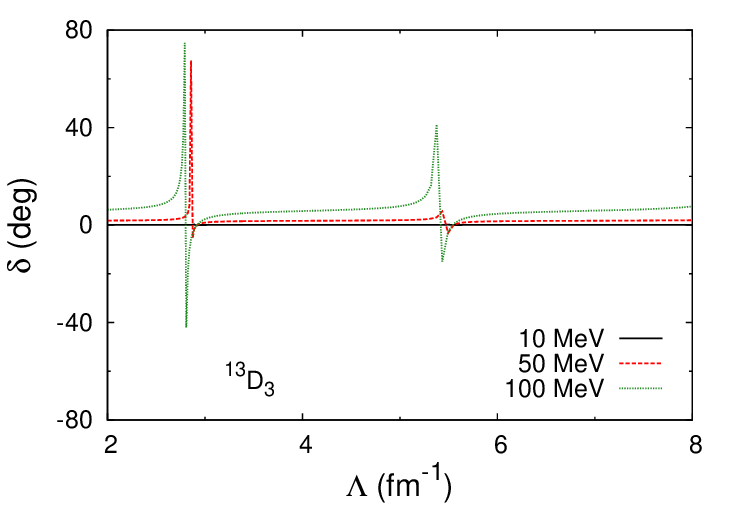} \hspace{2em}
	\includegraphics[width=0.45\textwidth]{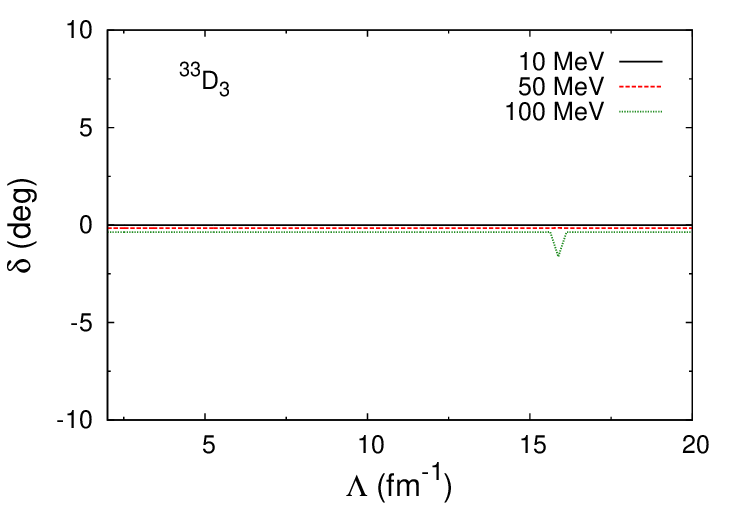} \\
	\includegraphics[width=0.45\textwidth]{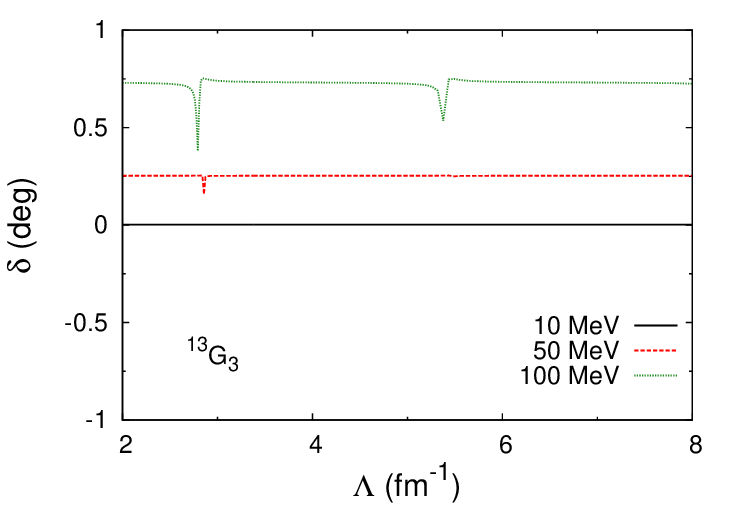}\hspace{2em}
	\includegraphics[width=0.45\textwidth]{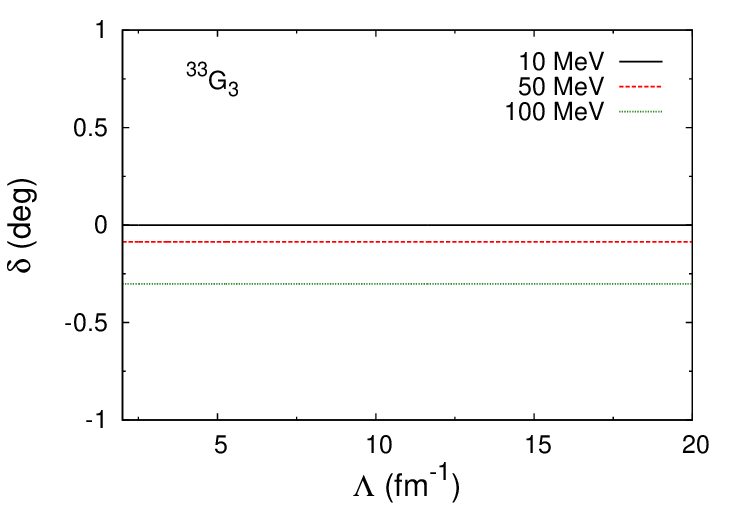}\\
	\includegraphics[width=0.45\textwidth]{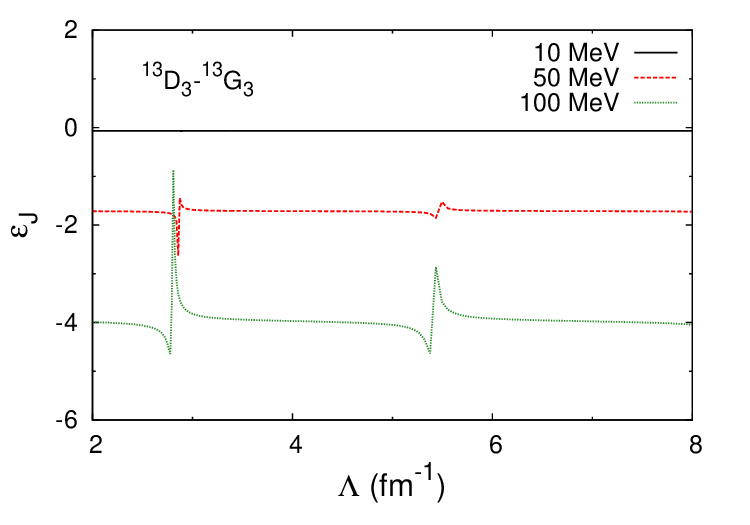} \hspace{2em}
	\includegraphics[width=0.45\textwidth]{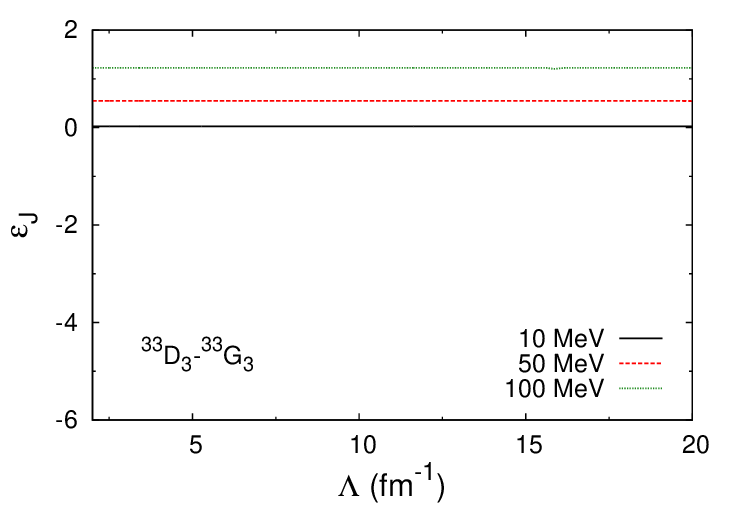}\\
	\caption{\label{TripletDGPhases_L}{Cutoff dependence of the phase shifts and mixing angles $\varepsilon_{J}$ in the spin-triplet coupled $D$-$G$ waves at the laboratory energies of 10 MeV (black solid line), 50 MeV (red dashed line), and 100 MeV (green dotted line), for $V_c=W_c=0~\text{fm}^{-1}$.}}
\end{figure}

\begin{figure}[tb]
	\centering
	\includegraphics[width=0.45\textwidth]{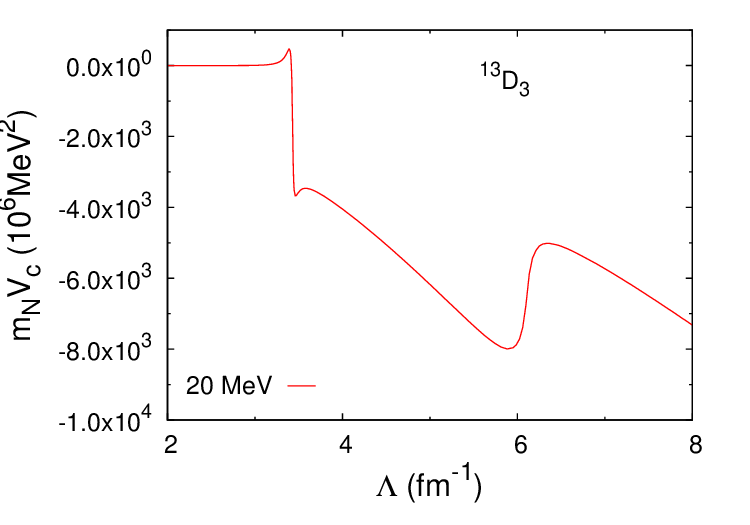} \hspace{2em}
	\includegraphics[width=0.45\textwidth]{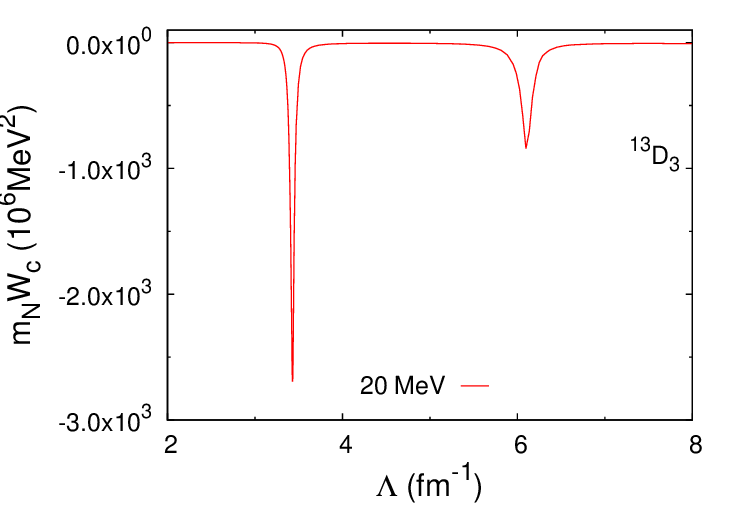} \\
	\includegraphics[width=0.45\textwidth]{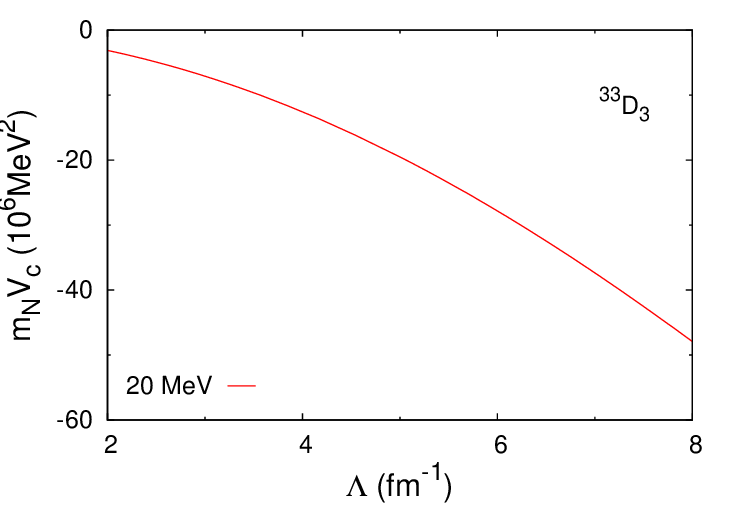} \hspace{2em}
	\includegraphics[width=0.45\textwidth]{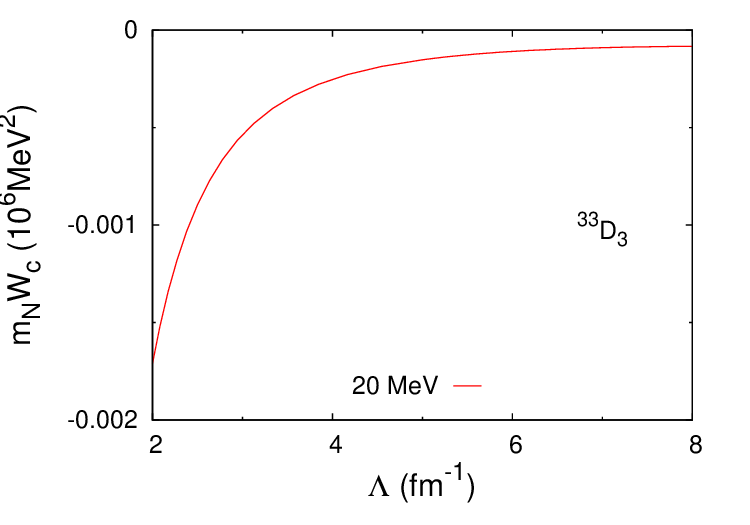} \\
	\caption{\label{TripletDDCount_L}{Cutoff dependence of $m_N V_c$ (left panels) and $m_N W_c$ (right panels) for the spin-triplet coupled $^{13}D_3$ and $^{33}D_3$ waves. The PWA phase shifts and inelasticities are fitted at $T_{\rm lab}=20$ MeV.}}
\end{figure}

Now let us look at the spin-triplet coupled partial waves,
which are the $^{13}D_3$-$^{13}G_3$ and $^{33}D_3$-$^{33}G_3$ waves in this paper.
The cutoff dependence of the phase shifts and their mixing angles $\varepsilon_J$
before renormalization are shown in Fig.~\ref{TripletDGPhases_L}.
Again the inelasticities $\eta$s are not shown since they all equal $1$ as $W_c=0~{\rm fm}^{-1}$ here.
The mixing angles $\omega_J$ are not shown either, because they are not well determined in this case
as one can see from Eq.~\eqref{Eq:Klarsfeld}.
One can find that those of the $^{13}D_3$-$^{13}G_3$ wave are obviously cutoff dependent at higher energies, while those of the $^{33}D_3$-$^{33}G_3$ wave are almost cutoff independent
except for the phase shift of $^{33}D_3$ at $T_{\rm lab}=100$ MeV.
Though both of these coupled waves have one eigenchannel which contains attractive singular tensor force, the cutoff-dependence behaviours are different for these two coupled waves. The difference shall be due to the different isospin factors.
Anyhow, these two coupled waves need contact interactions at LO to remove the cutoff dependence,
i.e., the contact interactions shall be promoted to LO though which are not there according to NDA.
Again, in Fig.~\ref{TripletDGPhases_L}, the cutoff ranges of the isospin-singlet and the isospin-triplet waves are different, because we also want to see the cutoff dependence in the isospin-triplet waves since that is expected from the experience in the lower partial waves
of $\overline{N}\!N$ system~\cite{Zhou:2022}.

\begin{figure}[tb]
	\centering
	\includegraphics[width=0.45\textwidth]{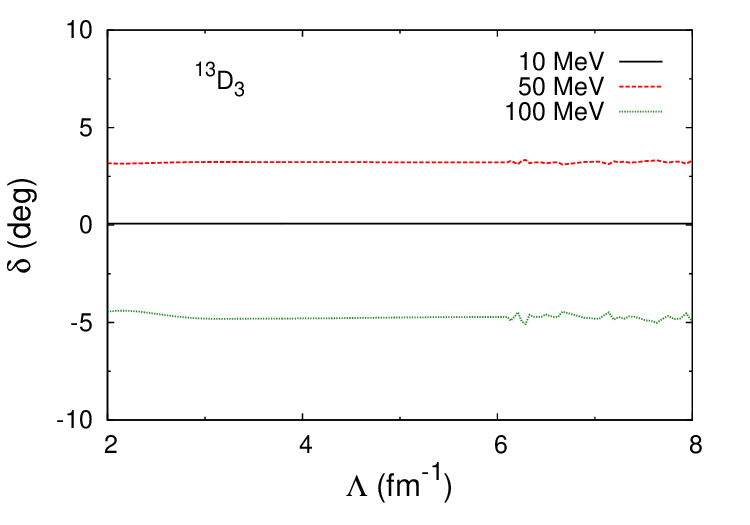} \hspace{2em}
	\includegraphics[width=0.45\textwidth]{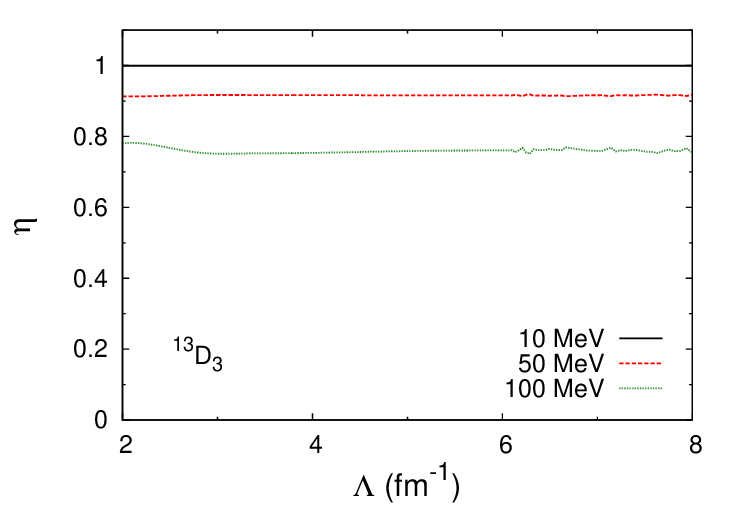}\\
	\includegraphics[width=0.45\textwidth]{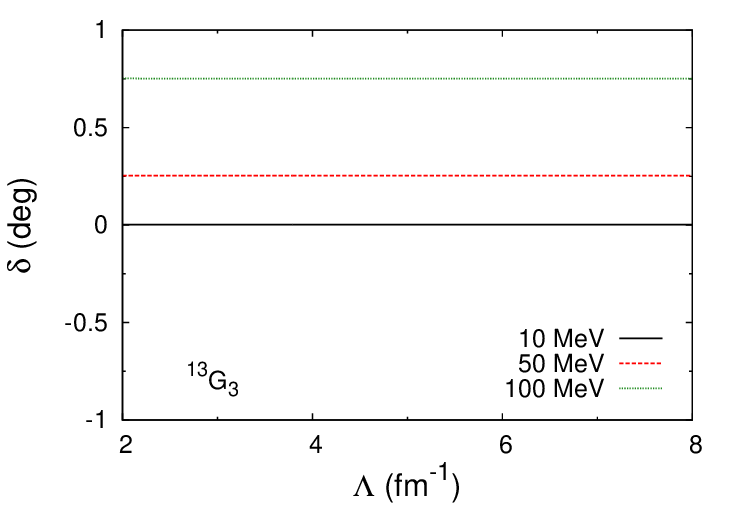} \hspace{2em}
	\includegraphics[width=0.45\textwidth]{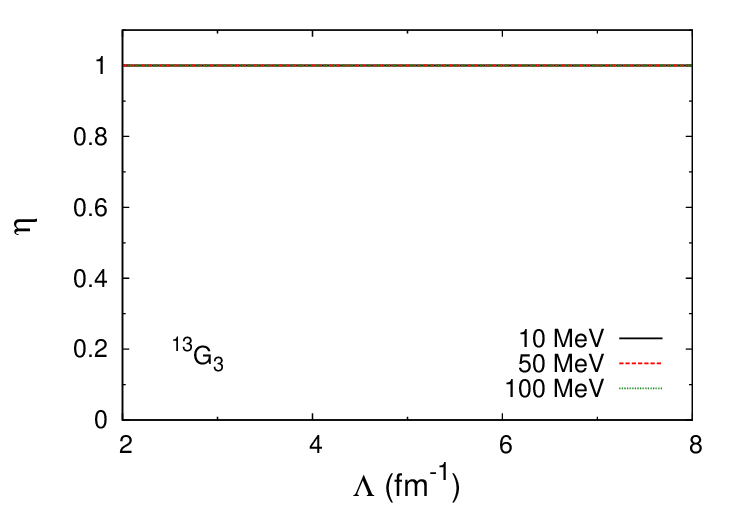} \\
	\includegraphics[width=0.45\textwidth]{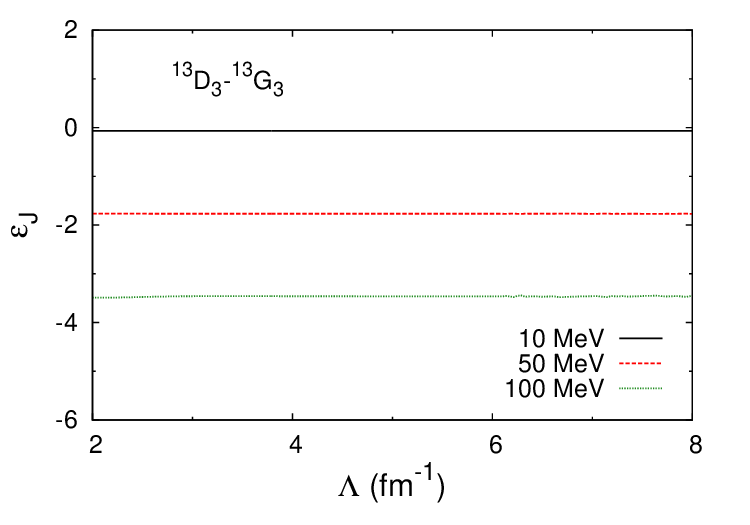} \hspace{2em}
	\includegraphics[width=0.45\textwidth]{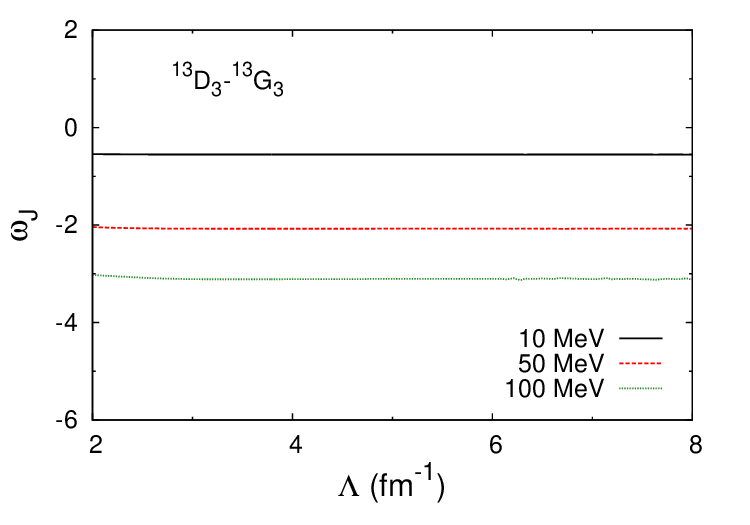}\\
	\caption{\label{Triplet1DGPhase_LC}{Residual cutoff dependence of the phase shifts, inelasticities, and mixing angles in the spin-triplet coupled $^{13}D_3$-$^{13}G_3$ waves at the laboratory energies of 10 MeV (black solid line), 50 MeV (red dashed line), and 100 MeV (green dotted line), for $V_c$ and $W_c$
	in Fig.~\ref{TripletDDCount_L}.}}
\end{figure}

\begin{figure}[tb]
	\centering
	\includegraphics[width=0.45\textwidth]{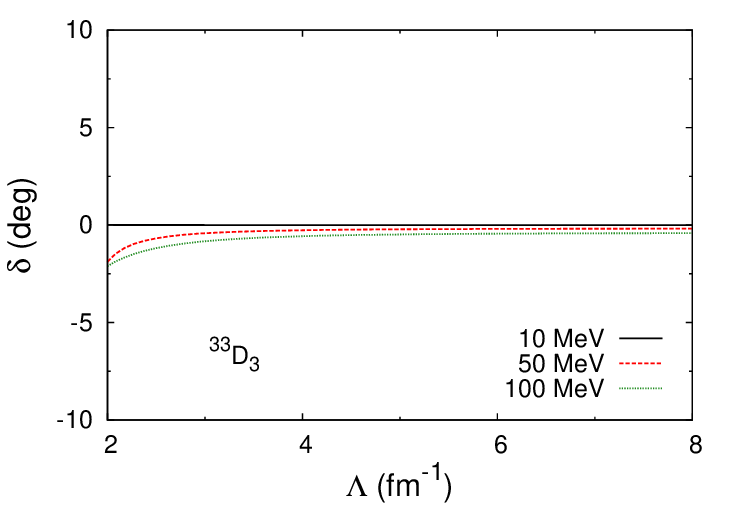} \hspace{2em}
	\includegraphics[width=0.45\textwidth]{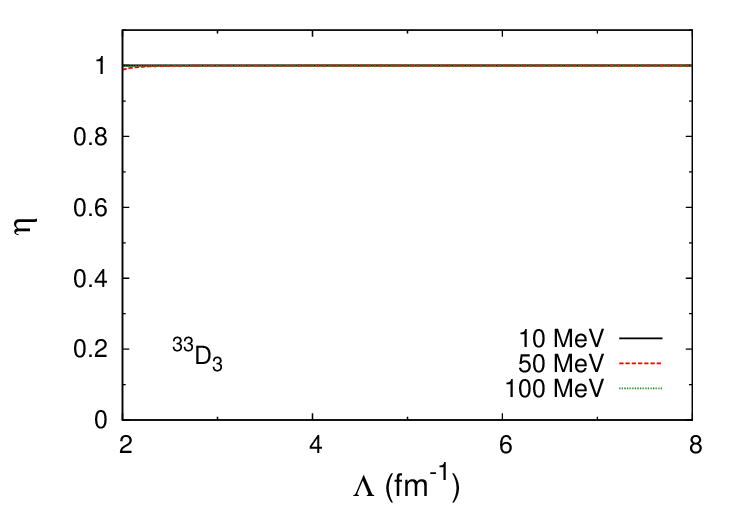}\\
	\includegraphics[width=0.45\textwidth]{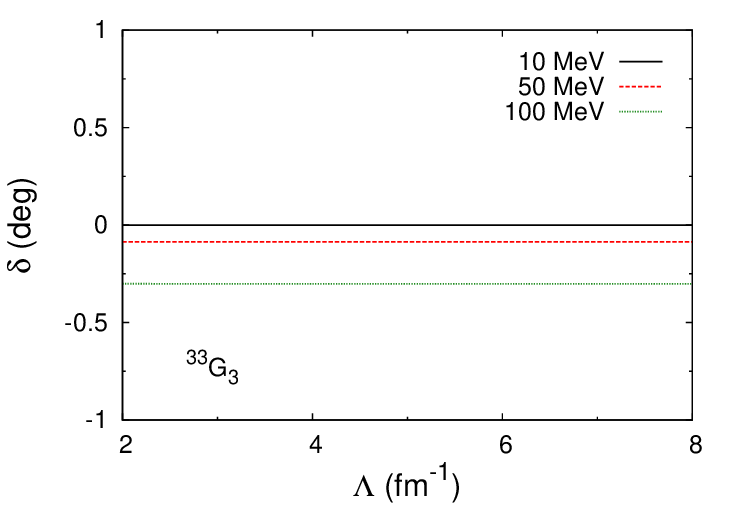} \hspace{2em}
	\includegraphics[width=0.45\textwidth]{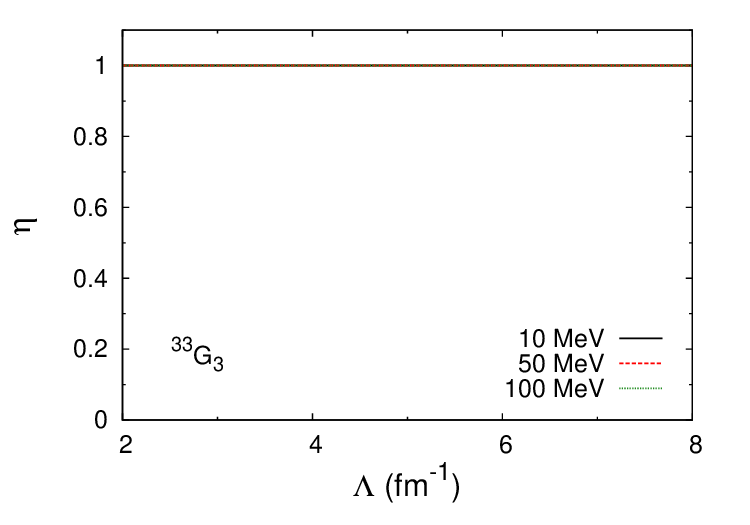} \\
	\includegraphics[width=0.45\textwidth]{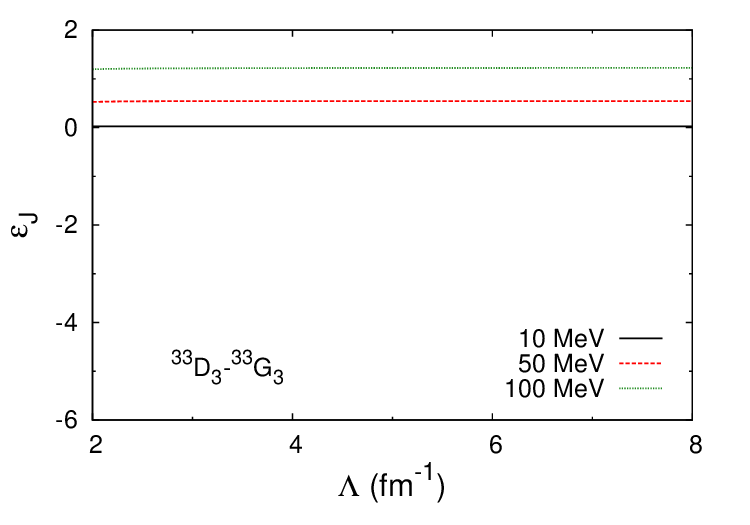} \hspace{2em}
	\includegraphics[width=0.45\textwidth]{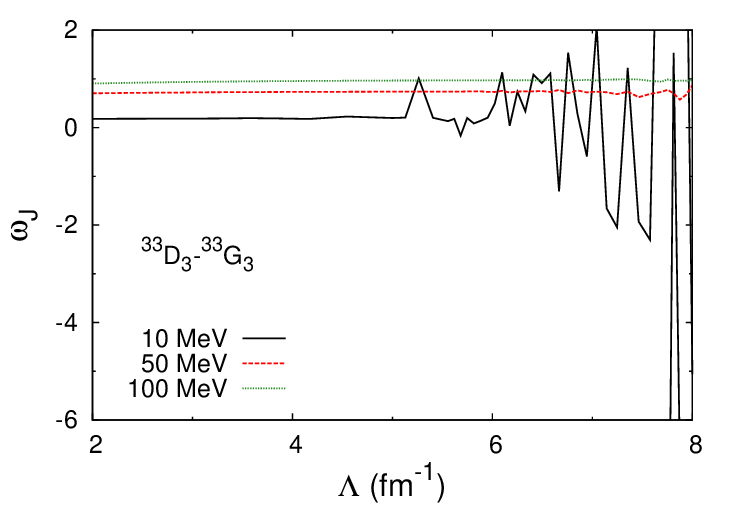}\\
	\caption{\label{Triplet3DGPhase_LC}{Residual cutoff dependence of the phase shifts, inelasticities, and mixing angles in the spin-triplet coupled $^{33}D_3$-$^{33}G_3$ waves at the laboratory energies of 10 MeV (black solid line), 50 MeV (red dashed line), and 100 MeV (green dotted line), for $V_c$ and $W_c$
	in Fig.~\ref{TripletDDCount_L}.}}
\end{figure}

The cutoff dependence of the LECs $V_c$ and $W_c$ of the $^{13}D_3$ and $^{33}D_3$ waves
are shown in Fig.~\ref{TripletDDCount_L}.
Again, the values of these LECs are obtained by fitting to the corresponding phase shifts and inelasticities of the PWA at laboratory kinetic energy $T_{\rm lab}=20$ MeV.
In general, the magnitude of $V_c$ is larger than the magnitude of corresponding $W_c$.
One can also see that the curves of $V_c$ and $W_c$ in the $^{13}D_3$ wave have more structures
than those in the $^{33}D_3$ wave, which are the reflections of Fig.~\ref{TripletDGPhases_L}.
The residual cutoff dependence of the phase shifts, inelasticities, and mixing angles
are shown in Figs.~\ref{Triplet1DGPhase_LC} and \ref{Triplet3DGPhase_LC}.
There are some ``ripple'' in the plots of $^{13}D_3$ phase shift and inelasticity, which are due to the difficulty of fitting in this wave when $b$ becomes smaller or equivalently $\Lambda$ becomes larger.
It looks like there is a weird oscillatory behaviour of the mixing angle $\omega_J$ for $^{33}D_3$-$^{33}G_3$
at $T_{\text lab}=10$ MeV in Fig.~\ref{Triplet3DGPhase_LC}.
However, this is not a problem because the inelasticities of $^{33}D_3$ and $^{33}G_3$ are both equal to $1$
in the cutoff range and thus $\omega_J$ can take any values in this case as can be seen from Eq.~\eqref{Eq:Klarsfeld}.
Therefore, one can see that the observables are cutoff independent after renormalization.
The values of $V_c$ and $W_c$ at $b=0.2$ fm for the coupled $D$-$G$ waves are given in Table~\ref{tab:Coup_vw},
which will be used later on.

\begin{table}[t]
	\centering
	\caption{Values of the real ($V_c$) and imaginary ($W_c$) components of the short-range potential at $b=0.2$ fm for the coupled $D$-$G$ partial waves. These of $^{3}D_3$ waves are obtained by fitting to the PWA ``data"~\cite{Zhou:2012ui} at $T_{\rm lab}=20$ MeV, while these of $^{3}G_3$ waves are set to be
	$0~{\rm fm}^{-1}$ by hand.}
	\tabcolsep=2.6em
	\renewcommand{\arraystretch}{0.9}
	\begin{tabular}{cd{7.2}d{1.0}d{2.0}d{1.0}}
		\hline
		\hline
		Partial wave &\multicolumn{1}{c}{$^{13}D_3$}&\multicolumn{1}{c}{$^{13}G_3$}
		            &\multicolumn{1}{c}{$^{33}D_3$}&\multicolumn{1}{c}{$^{33}G_3$}\\
		\hline
		$V_c$ (fm$^{-1}$) & -33336.83 & 0 & -105.36 & 0 \\
		$W_c$ (fm$^{-1}$) & -31.18 & 0 & -0.00 & 0 \\
		\hline
		\hline
	\end{tabular}
	\label{tab:Coup_vw}
\end{table}

\begin{figure}[tb]
	\centering
	\includegraphics[width=0.45\textwidth]{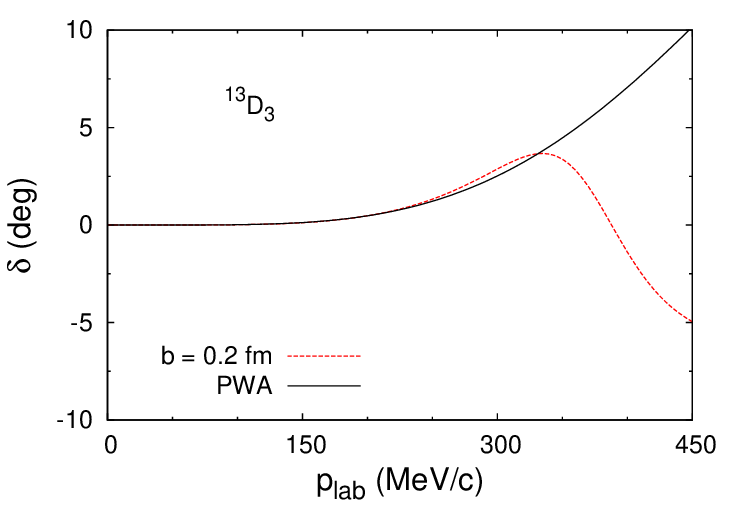} \hspace{2em}
	\includegraphics[width=0.45\textwidth]{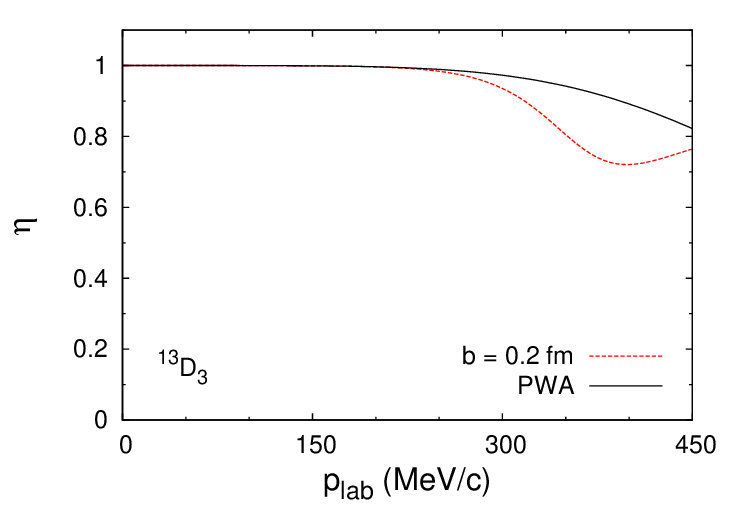} \\
	\includegraphics[width=0.45\textwidth]{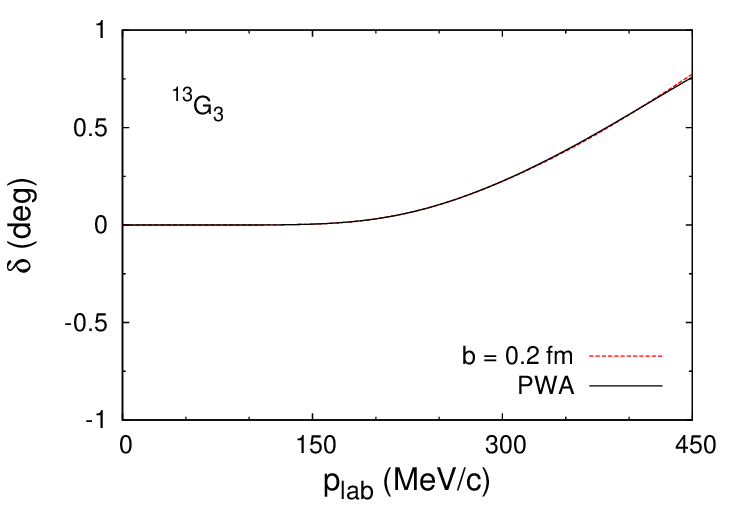} \hspace{2em}
	\includegraphics[width=0.45\textwidth]{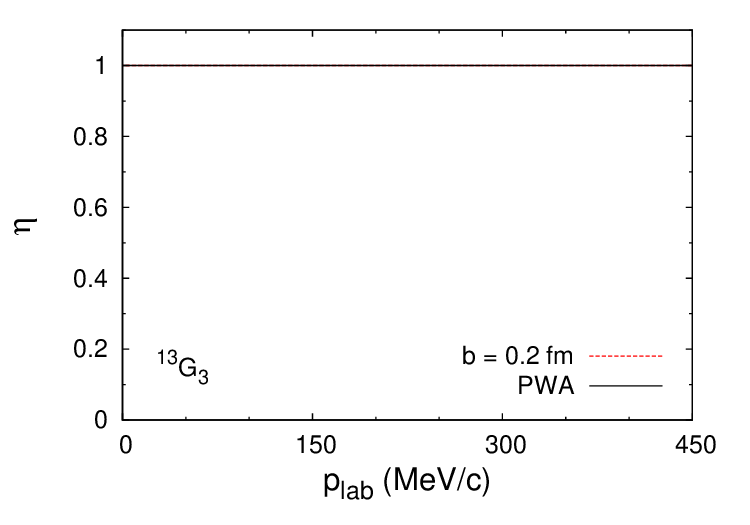}\\
	\includegraphics[width=0.45\textwidth]{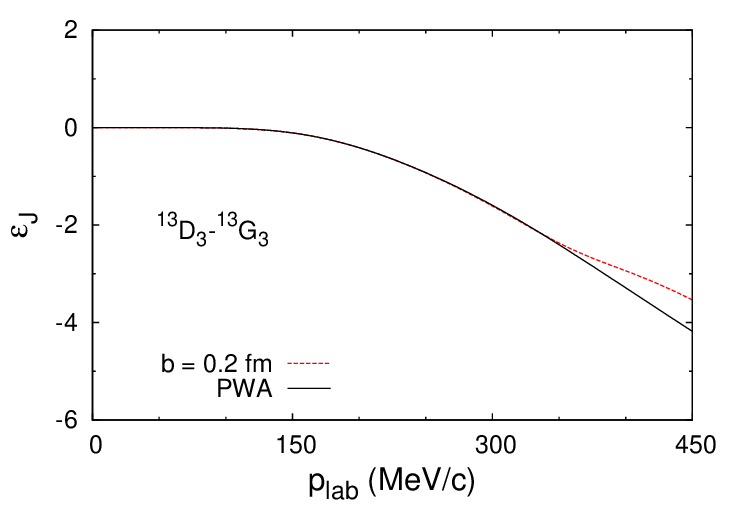} \hspace{2em}
	\includegraphics[width=0.45\textwidth]{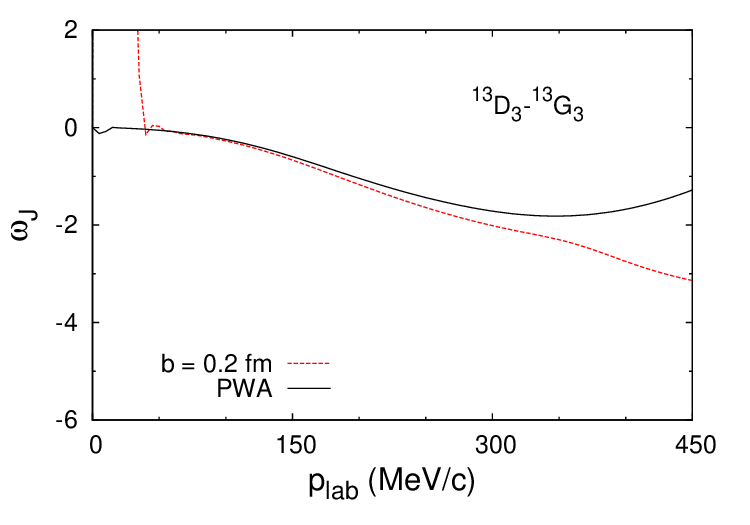}\\
	\caption{\label{Triplet1DGPhase_plab}{Phase shifts and mixing angle (left panels) and inelasticities and their mixing angle (right panels) of the spin-triplet coupled $^{13}D_3$-$^{13}G_3$ waves against laboratory momentum. The (red) dashed lines are from iterated one-pion exchange for $b=0.2$ fm
	and $V_c$, $W_c$ from Table~\ref{tab:Coup_vw}, while (black) solid lines are the results of the PWA~\cite{Zhou:2012ui}.}}
\end{figure}

\begin{figure}[tb]
	\centering
	\includegraphics[width=0.45\textwidth]{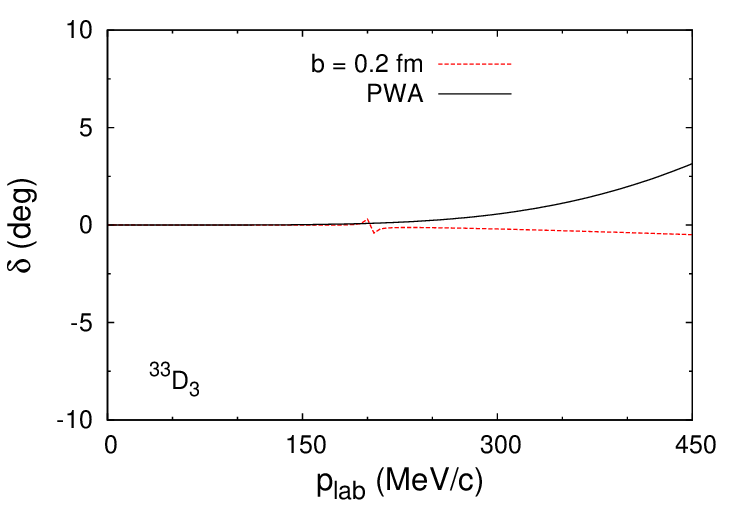} \hspace{2em}
	\includegraphics[width=0.45\textwidth]{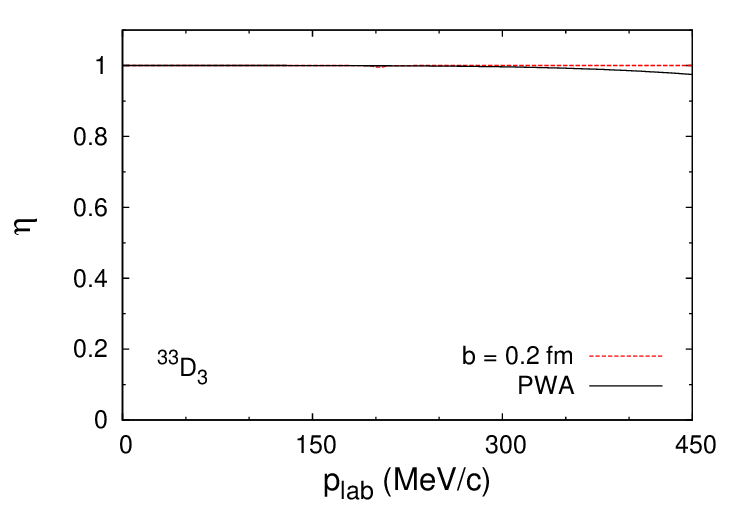} \\
	\includegraphics[width=0.45\textwidth]{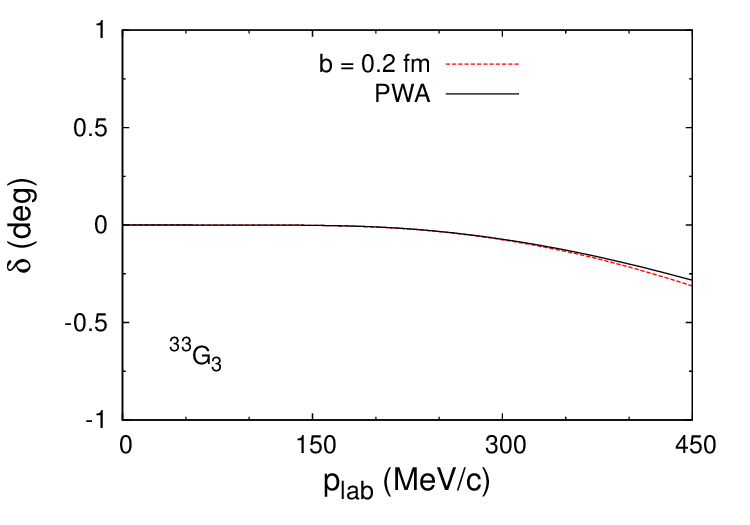} \hspace{2em}
	\includegraphics[width=0.45\textwidth]{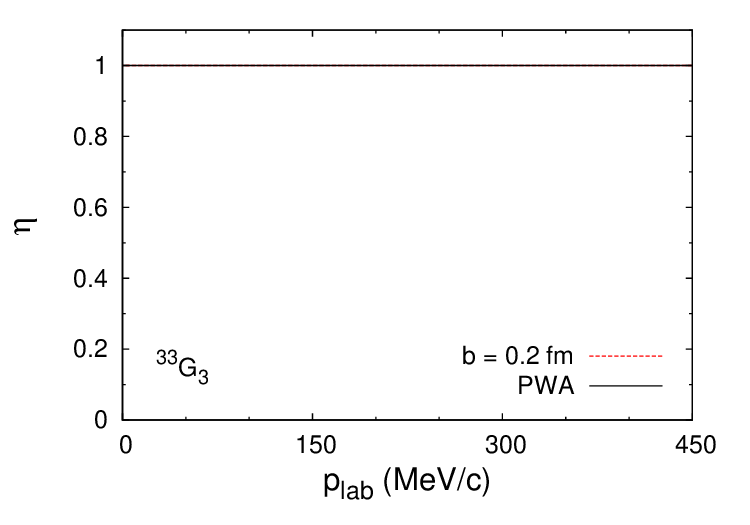}\\
	\includegraphics[width=0.45\textwidth]{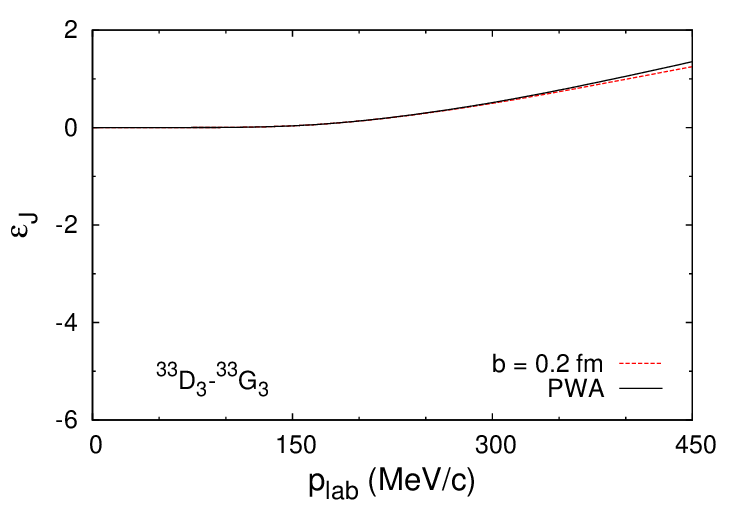} \hspace{2em}
	\includegraphics[width=0.45\textwidth]{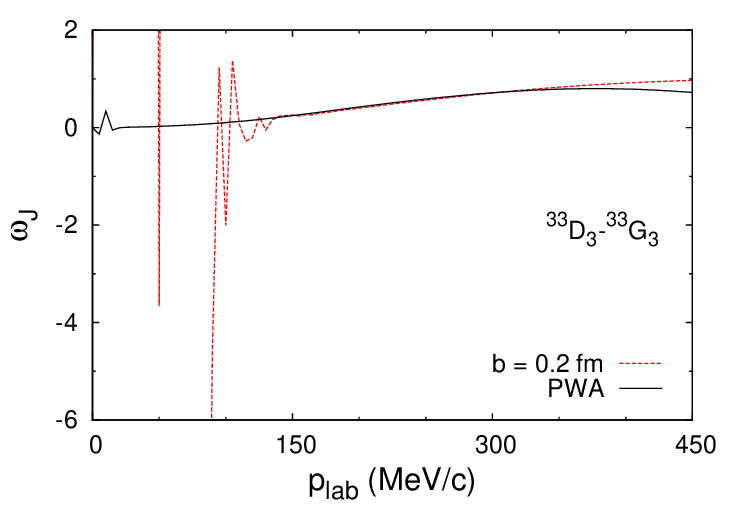}\\
	\caption{\label{Triplet3DGPhase_plab}{Phase shifts and mixing angle (left panels) and inelasticities and their mixing angle (right panels) of the spin-triplet coupled $^{33}D_3$-$^{33}G_3$ waves against laboratory momentum. The (red) dashed lines are from iterated one-pion exchange for $b=0.2$ fm
	and $V_c$, $W_c$ from Table~\ref{tab:Coup_vw}, while (black) solid lines are the results of the PWA~\cite{Zhou:2012ui}.}}
\end{figure}

After renormalization, the phase shifts, inelasticities and mixing angles against laboratory momentum, and the comparison with the PWA values are shown in Figs.~\ref{Triplet1DGPhase_plab} and \ref{Triplet3DGPhase_plab}.
The values of $V_c$ and $W_c$ for the iterated OPE are taken from Table~\ref{tab:Coup_vw}.
The results agree with the PWA very well except for the phase shift and inelasticity of $^{13}D_3$
when $p_{\rm lab} \simge 300$ MeV/c or so, and the phase shift of $^{33}D_3$
when $p_{\rm lab} \simge 200$ MeV/c or so.
There is also a dip-bump structure in the (red) dashed phase-shift curve of $^{33}D_3$
around $p_{\rm lab}=200$ MeV/c in Fig.~\ref{Triplet3DGPhase_plab}.
All of these discrepancies might be hints that we should treat OPE perturbatively in these waves.
Of course, this need further investigation.
It seems that the (red) dashed lines in Figs.~\ref{Triplet1DGPhase_plab} and \ref{Triplet3DGPhase_plab} of the mixing angles $\omega_J$ have weird behaviours in the low energy range.
Again, these do not matter. This is because of that the inelasticities are very close or equal to $1$ at these energies, thus $\omega_J$ is not well determined as can be seen from Eq.~\eqref{Eq:Klarsfeld}. In other words, $\omega_J$ can take any values in this case.
The very small dip-bump structures of the PWA curves of $\omega_J$ in Figs.~\ref{Triplet1DGPhase_plab}
and \ref{Triplet3DGPhase_plab} may also be due to the same reason.
The quality of the agreement with the PWA of the coupled $D$-$G$ waves is comparable with
that in the $N\!N$ system~\cite{Nogga:2005hy}.

\section{Conclusions and outlooks} \label{sec:Conclusions}

In summary, as an extension of the previous work \cite{Zhou:2022},
we have adapted the framework of the partial-wave analysis
for $\overline{N}\!N$ scattering \cite{Zhou:2012ui}
to study the renormalization of the iterated static OPE potential in
the other partial waves with total angular momentum $J\le3$,
which are the uncoupled $D$, $F$ and coupled $D$-$G $ waves.

For the spin-singlet channels $^{11}D_2$, $^{31}D_2$, $^{11}F_3$ and $^{31}F_3$,
they have no tensor forces.
We have seen that their phase shifts are cutoff independent when $V_c=W_c=0~{\rm fm}^{-1}$,
therefore no counterterms are needed in these spin-singlet channels.
For the spin-triplet uncoupled channels, we have confirmed that the renormalization of iterated OPE requires
counterterms in the $^{33}D_2$ and $^{33}F_3$ channels
which have attractive singular tensor forces,
while the $^{13}D_2$, $^{13}F_3$ channels do not need counterterms since they have repulsive singular tensor forces.
For the spin-triplet coupled channels, we have confirmed that $^{13}D_3$ and $^{33}D_3$ need counterterms
while $^{13}G_3$ and $^{33}G_3$ do not, because one eigenchannel is attractive and the other is repulsive.

According to NDA, the counterterms start to appear on the order of $Q^4$
for the $^{33}D_2$ channel,
and start to appear on the order of $Q^6$
for the $^{33}F_3$ and $^{3}D_3$-$^{3}G_3$ channels.
The RG invariant requires that these counterterms need to be promoted to LO as we have seen.
After renormalization, most of the observables agree with the PWA values very well
in the energy range considered,
except that some of them agree with the PWA values well in some lower energy ranges.
The discrepancies between the iterated OPE results and the PWA values might be reduced
when high order contributions, such as TPE, were included.
Of course, this need further investigation,
because TPE is treated perturbatively as required by renormalization,
and it is claimed in Ref.~\cite{Gasparyan:2022}
that a cutoff independent scattering amplitude can in general not be obtained in the infinite-cutoff (RG-invariant) scheme
beyond the leading order in spite of the perturbative treatment of subleading contributions.

The magnitudes of $V_c$ are much greater than those of the corresponding $W_c$ as can be seen from
Figs.~\ref{TripletUncpCount_L} and \ref{TripletDDCount_L}.
This is already the case in the lower partial waves of $\overline{N}\!N$ system~\cite{Zhou:2022},
but it is even more so here.
The $W_c$ in the $^{33}D_2$, $^{33}F_3$ and $^{33}D_3$ waves might even be set to be $0~{\rm fm}^{-1}$,
because they are relatively so small as comparing to the corresponding $V_c$.
In the $^{13}D_3$ wave, there are some sharp spikes in the very limited cutoff ranges.
The magnitudes of these spikes are comparable with the magnitude of the corresponding $V_c$.
Therefore, it seems that in some higher partial waves,
the imaginary part of an LEC is not on the same order as the corresponding real part lies.
We leave this issue as a future study.

As we have seen that the results are satisfactory when OPE is treated nonperturbatively.
However, it might be unnecessary to treat OPE nonperturbatively in some high partial waves.
In most of these high partial waves, the perturbative results should be similar to the nonperturbative ones
owing to the smallness of the phase shifts.
In a future investigation, we will attempt to study the renormalization in the $\overline{N}\!N$ system
where OPE will be treated perturbatively,
and try to identify the boundary between nonperturbation and perturbation in OPE.

\FloatBarrier

\section*{Acknowledgments}
The author thank R. G. E. Timmermans, U. van Kolck and B. Long for useful discussions and suggestions.
This work was supported by the Doctoral Fund Project under grant No.~2020BQ03 of Nanfang College, Guangzhou.

\end{document}